\documentclass[12pt]{article}
\usepackage{amsmath, amssymb, graphics}
\usepackage{amsfonts}
\usepackage{amsmath}
\usepackage{graphicx}

\newcommand{\mathsym}[1]{{}}

\def\IR{\mathbb{R}}

\def\id{\protect{{1 \kern-.28em {\rm l}}}}

\def\be{\begin{eqnarray}}
\def\ee{\end{eqnarray}}

\makeatletter
\renewcommand\section{\@startsection {section}{1}{\z@}%
                                   {-3.5ex \@plus -1ex \@minus -.2ex}%
                                   {2.3ex \@plus.2ex}%
                                   {\normalfont\large\bfseries}}
\renewcommand\subsection{\@startsection{subsection}{2}{\z@}%
                                   {-3.25ex\@plus -1ex \@minus -.2ex}%
                                   {1.5ex \@plus .2ex}%
                                   {\normalfont\normalsize\bfseries}}
\makeatother

\def\b{{\rm b}} %$\beta$ in KRTT
\def\Tr{{\rm Tr}}

\def \foot {\footnote}
\def \bi{\bibitem}

\def \ha {{1 \over 2}}
\def \td {\tilde}
\def \ci{\cite}
\def \N {{\mathcal N}}

\def\S{{\mathcal S} }

\def \E {{\mathcal  E}} \def \J {{\mathcal  J}}

\def\b{\beta}

\def\Tr{{\rm  Tr}}

\def \del{\partial}

\def\s{\sigma}

\def\ov{\over}

\def\J{{\mathcal J}}

\def\E{{\mathcal E}}

\def\b{\beta}
\def\l{\lambda}

\def \k {\kappa}
\def\foot{\footnote}

\def \ci {\cite}

\def \foot {\footnote}
\def \bi{\bibitem}

\def \ha {{1 \over 2}}

\def \ep {\epsilon}

\def \Tr {{\rm Tr}}

\def \l  {\lambda}

\def \N {{\mathcal N}}
\def \S {{\rm S}}

\def \td {\tilde}

\def \N {{\mathcal N}}

\def \bi{\bibitem}
\def \la {\label}

\def \l {\lambda}
\def\foot{\footnote}

\def \sql {{\sqrt \l}}
\def \adss {$AdS_5 \times S^5~$ }
\newcommand{\rf}[1]{(\ref{#1})}
\def \ov {\over}

\def\N{{\cal N}}

\def \ha{{1\ov 2}}

\def\r{{\rm r}}

\def \no {\nonumber}

\def \J {\mathcal{J}}
\def \del {\partial}

\def \E {{\cal E}}

\def \S {{\cal S}}
\def \J {{\cal J}}

\def\ms{\mathcal{S}}

\def \bi{\bibitem}
\def \la {\label}

\def \l {\lambda}
\def\foot{\footnote}

\def \sql {{\sqrt \l}}

\def \adss {$AdS_5 \times S^5$\ }

 \def \r {\rho}

\def \bu {{\bar u}}

\def \ov {\over}

\def \varpi {{\rm w}}

\def \ep {\epsilon}

\def\pic #1#2{\hbox{\lower#1pt\hbox{~\mbox{\epsfxsize=20truemm \epsffile{#2}}}}}

\def\pic #1#2#3{\hbox{\lower#1pt\hbox{~\mbox{\includegraphics[scale=#3]{#2}}}}}

\def\IR{\mathbb{R}}

\def\be{\begin{eqnarray}}
\def\ee{\end{eqnarray}}

\def\id{\protect{{1 \kern-.28em {\rm l}}}}

\def \S  {{\rm S}}

\def \mm {{\cal  \ell}}

\def \be {\bea}
\def \ee {\eea}

\def\ff{~{{\tilde \phi}}}

\def\Tr{{\rm Tr}}

\def\IR{\mathbb{R}}

\def\id{\protect{{1 \kern-.28em {\rm l}}}}

\def\be{\begin{eqnarray}}
\def\ee{\end{eqnarray}}

\def\b{{\rm b}} %$\beta$ in KRTT
\def \bea { \be}
\def \eea {\ee}

\def \b {\beta}
\def \bs {\bigskip}

\def \del {\partial} 

\def \s {\sigma}

 \def \J {{\cal J}}
 \def \S {{\cal S}}
 \def \E {{\cal E}}
\def \fix {{ \rm fixed}}

\def \jj  {\ell} 
\def \ff  {{\rm f}}

\def \tT {{_{\rm tot}}}
\def\jjj{{\ell}}

\def \ws {{\rm 2d}}

\def \st {{ }}

\def \tT {{\cal T}}
\def \rT {{\rm T}}

\def \lnu {{{\hat  \ell}\,}}
\def \tE  {{\rm f}}
\def\hkappa{{\hat\kappa}}

\def\cM{{\cal M}}

\def \rmh {{\rm m}}
\def \Sigm {{\cal D}}
\def \hu {\hat u}
\def \bu {u}
\def \G {\Gamma} 
\def \ms {\medskip}
\def \mm {{m}}
\def \kk {{\rm k}}
\def \bmm {{\rm m}}

\begin{document}

%%%%%%%%%%%%%%%%%%%%%%%%%%%%%%%%%%%%%%

\overfullrule=0pt
\parskip=2pt
\parindent=12pt
\headheight=0in \headsep=0in \topmargin=0in \oddsidemargin=0in

\vspace{ -3cm}
\thispagestyle{empty}
\vspace{-1cm}

\rightline{Imperial-TP-AT-2007-5}

\rightline{NI07095}

\begin{center}
\vspace{1cm}
{\Large\bf  

%Charged
 Spinning superstrings  at two loops: \\
 strong-coupling corrections to 
 dimensions \\
 of large-twist SYM operators

\vspace{1.2cm}

\vspace{0.3cm}

   }

\vspace{.2cm} {
R. Roiban$^{a,b,}$\footnote{radu@phys.psu.edu} and A.A. Tseytlin$^{b,c,}
$\footnote{Also at Lebedev  Institute, Moscow. tseytlin@imperial.ac.uk }}\\

\vskip 0.6cm

{\em 
$^{a}$Department of Physics, The Pennsylvania  State University,\\
University Park, PA 16802 , USA\\
\vskip 0.08cm
$^b$The Isaac Newton Institute, Cambridge, U.K.
 \\
\vskip 0.08cm $^{c}$Blackett Laboratory, Imperial College,
London SW7 2AZ, U.K. }

\end{center}

\begin{abstract}
 %%%%%%%%%%%%%%%%%%%%%%%%%%%%%%%%%
 We consider folded $(S,J)$ spinning strings in $AdS_5 \times S^5$   (with 
 one spin  component in $AdS_5$ and a  one in $S^5$) corresponding to the 
 Tr$(D^S \Phi^J)$  operators in the $sl(2)$ sector of the $\N=4$ SYM theory in
 the special  scaling limit in which both the string mass 
  $\sim \sqrt \lambda \ln S$  and $J$  are  sent  to infinity 
 with  their ratio  fixed. Expanding in the parameter $\ell= {J \ov \sqrt
  \lambda \ln S}$
 we compute the 2-loop string sigma model correction to the string energy
 and show that it 
 agrees with the expression proposed by Alday and Maldacena
 in arxiv:0708.0672. 
 We suggest that a resummation of the  logarithmic $\ell^2 \ln^n \ell$ 
 terms is necessary  in order to establish  an interpolation to the weakly 
 coupled gauge theory
 results. In the process, we  set up a general  framework for the calculation 
 of higher loop corrections 
to the energy of multi-spin  string configurations. In particular, 
we find that in addition to the  direct 2-loop term in  the  string energy  there
is a contribution from lower loop order  due to a finite ``renormalization'' 
of the relation between the parameters  of the classical solution and the  
fixed spins, i.e. 
the charges   of the $SO(2,4) \times SO(6)$ symmetry.

%%%%%%%%%%%%%%%%%%%%%%%%%%%%%%%%%
\end{abstract}

\newpage

\setcounter{equation}{0} 
\setcounter{footnote}{0}
\setcounter{section}{0}

%%%%%%%%%%%%%%%%%%%%%%%%%%%%%%%%%%%%%%

\section{Introduction \label{intro}}
%%%%%%%%%%%%%%%%%%%%%%%%%%%%%%%%%%%
The  spinning folded closed string state
in $AdS_5$   for which the  difference between the energy $E$ and the spin $S$ 
scales as $\ln S$ 
\ci{gkp}  played a remarkable role  in the recent progress in the 
quantitative understanding of the AdS/CFT  duality (see, e.g., 
\ci{ft1,bfst,kot,bgk,be,bern,bes,ben,ftt,ald,krj,rtt,am2,bkk,rt}). 
With spin $J$ in $S^5$  added \ci{ft1},
  this state  can be thought of as being
   dual 
to Tr$(D^S \Phi^J)$ 
operators  in the $sl(2)$ sector  of the $\N=4$ SYM theory,  interpolating  between 
the near-BMN operators for $ J \gg S$ and small-twist operators for small $J$. 
The resulting  quantum string energy  $E(S,J, \sql)$ (or the  gauge theory anomalous dimension 
$\Delta=E-S-J$) is a  non-trivial function of three arguments that can be explored 
in various 
limits, uncovering and testing important features of the underlying 
Bethe ansatz \ci{bes}.

Our aim here   will to compute the 2-loop  string correction 
to this energy in  an important $J$-dependent scaling limit \ci{bgk,ftt},  
extending  earlier 2-loop result  found in the $J=0$ case \ci{rtt,rt},
%AT
and to compare it to the prediction made recently in \ci{am2} 
on the basis of a conjectured relation  of this scaling limit 
to  the $O(6)$ sigma model. 

\bs 

Let us begin by reviewing what is 
  known  about $E(S,J, \sql)$ in various relevant limits. 
String semiclassical expansion  can be  organized 
as an expansion  in  the inverse tension or $ 1\ov \sqrt{\lambda}$  with 
 the semiclassical parameters 
$\S={ S \ov  \sql}, \      \J=   { J \ov  \sql}$
(or ``frequencies'') 
being kept  fixed  
  \ci{ft1,ftt}
\be 
E &=& \sql\   \E ( \S,\J, { 1 \ov \sql}) =
\sql \bigg[  \E_0  ( \S,\J) +  { 1 \ov \sql}  \E_1  
( \S,\J)  + { 1 \ov (\sql)^2 } \E_2  ( \S,\J)  + ...
\bigg]\ . \la{ghi}
\ee
The semiclassical (SC)
string limit    is thus defined by 
\be
{\rm SC:} \ \ \ \ \ 
 \S,\ \J= {\rm fixed} \ , \ \ \ \ \l\to \infty \ ,\ \ \ \ \ \ \ \ \ \ 
 \S\equiv   { S \ov  \sql} \ ,  \ \ \     \J\equiv   { J \ov  \sql} 
  \ . 
\la{jk} 
\ee
%
%   On the gauge theory side 
%   one usually fixes (large) values of   $S$ and $J$
%   and expands in  $\l$ % to be  small or large
%   independently of $S,J$. 
%
 The semiclassical  string expansion thus explores  the  energy/dimension  
 in the 3-parameter space 
 $(S,J,\l)$  far away  from the origin  along the  ``diagonals''   with  
 $S \sim \sql$ and $J \sim \sql$.  
 This should be kept in mind when comparing to gauge theory, where 
 one usually fixes the (large) values of   $S$ and $J$
  and  studies the dependence on   $\l$ % to be  small or large
  for fixed  $S,J$. 
 
Within this semiclassical limit we  may consider a special ``sub-limit'' \ci{bgk}
which we shall   call the  ``semiclassical scaling limit'' or SCS  (also called
the ``long string limit'' in \ci{ftt}) 
% \ci{bgk,ftt}
\be 
{\rm SCS:} \ \ \  \ \ \    \S \gg \J \gg  1  \ , \ \ \  \ \ \ \ \ \   
\jj\equiv  
{ \J \ov {1 \ov \pi} \ln \S } = \fix   \ .   \la{ho}
\ee 
Since  $\ln \S \gg \ln \J$  and $S \gg \sql$ 
  we may as well  assume
%,
%N
% as 
%was done in   \ci{bgk,ftt},
 that this limit   is defined  by 
\be  \la{fi}
{\rm SCS:} \ \ \  \ \ \  S \gg J \gg  1 \ , \ \ \ \ \ \ \ \ 
 \jj\approx  
 % { \J \ov  {1 \ov \pi} \ln {\S\ov \J} }=  
   { J \ov  {\sql\ov \pi} \ln { S} }\equiv { j \ov \sql }= \fix  \ . \ee
Next, we  may also consider  ``sub-limits'' 
in which  the fixed parameter  ${ \J \ov \ln \S }$ 
or  ${ \J \ov \ln {S} }$ 
is  much smaller  (or much larger) 
 than 1 so that one can expand in powers of  it (or of its inverse) \ci{ftt}.
 %\foot{
 %
 % R
 %
%%Alternatively, one may consider a formal expansion
 %in positive or negative powers of $\ell$, without 
% assuming it to be either small or large.
% }
 Here we shall focus on the first possibility, i.e. on the 
 ``semiclassical scaling small'' 
limit or SCSS  (called ``slow long string limit''
%AAT
 in \ci{ftt})\foot{The parameter 
$\jj$ is the inverse of the parameter $x$ in  \ci{ftt} and is equal to $y$ 
in \cite{am2}. Our parameter $j=\sql \ell$ is related by a factor of $2\pi$
to the parameter $j$ in  \ci{am2}.}
\be \la{gh}
%AT
{\rm SCSS:} \ \ \ \ \ \ \ \ \ \ \  \jj =   
{ \J \ov { 1 \ov \pi}  \ln {\S} } \ \ll \ 1   \ . 
\ee 
 Note that the  SCS and SCSS   limits are  just the special cases 
 of the semiclassical limit \rf{jk} 
  limit, so  that the expansion \rf{ghi}  still applies, and  each term 
 $\E_n$ in  \rf{ghi}  can   be simplified   further  by taking these limits. 

\ms 

%NNEW

Motivated by the structure of the   mass  spectrum \ci{ftt} 
of the small fluctuations 
near the long $(S,J)$  string solution  in the scaling limit \rf{ho} 
it was  suggested in \ci{am2} that certain part of  the expression 
for the string energy in \rf{ghi}  can be  exactly  captured by the quantum 
$O(6)$  sigma model  provided  one considers the following limit:
\be \la{amm}
{\rm AM:} \ \ \ \ \ \ \ 
\l \gg 1 \ ,  \ \   S \gg J \gg 1 \ , \ \ 
\ \ \ \ \  j= { \pi J \ov \ln S}  \to 0 \ , 
\ \ \ \ \  \ \    
{ j \ov \mm  } \equiv  { \jj \ov \bmm } =  {\rm fixed}    \ .  
\ee 
Here $\mm$ is the dynamically generated scale of the $O(6)$ model 
\be \la{scm} 
\mm \approx \kk\ \l^{1/8}\  e^{- {1 \ov 4} \sql} \ \ll 1  \ , \ \ \ \ \ \ \ 
\bmm \equiv  { \mm \ov \sql} \approx \kk\ \l^{-3/8}\  e^{- {1 \ov 4} \sql}\  \ll 1  \ ,  \ee 
where $\kk$  is determined 
by how the $O(6)$ model is embedded into the \adss string theory 
(comparison  with the 1-loop superstring result of \ci{ftt} fixes 
it to be \ci{am2} $\kk= {2^{1/4}\ov \Gamma(5/4)}$). 
The part of the string energy that can be captured by the $O(6)$ model in this limit 
can be written as 
\be \la{ame}
{\rm AM:} \ \ \ \ \ \ \ \ \ \ \ \ \ \ \ \ \ 
E = {\textstyle{1 \ov \sql}}\  j^2\ \bar \E ({j\ov \mm }, {\textstyle{1 \ov \sql}} ) 
= \sql \ \ell^2  \  \bar \E ({\ell\ov \bmm },  {\textstyle{1 \ov \sql}})  \ . \ee
This  limit  in which $\ell$ scales to zero in a way correlated with the limit of 
$\lambda \to \infty$  as 
\be \la{cob}
 \ell\  \sim \  \bmm \   \sim \  \l^{-3/8}\  e^{- {1 \ov 4} \sql} 
  \ , \ee 
may be viewed as a {\it special  case}  of the SCSS limit \rf{gh}, i.e. of the 
limit   we shall consider below.

\bs

The $(S,J)$ string solution, given in general in terms of elliptic functions, 
simplifies dramatically 
in the scaling limit \rf{ho} \ci{ftt}: it becomes the following 
{\it homogeneous} configuration in  $AdS_3\times S^1$
\be
&& ds^2=-\cosh^2 \rho\ dt^2+d\rho^2+\sinh^2 \rho\  d\theta^2+d\phi^2\ , \no \\
&& 
t=\kappa\tau~,~~~~~~~~~\rho=\mu\sigma \ , \ \ \ \ \ \     \    \ \theta=\kappa\tau~,~~~~~~\phi=\nu
\tau~,\ \ \ \ \ \ \   \k,\mu,\nu  \gg 1   \ , 
\label{solc}
\ee
where  the conformal gauge condition requires that
\be
\kappa^2=\mu^2+\nu^2 \ . 
\label{Vir}
\ee
Here $0 \leq \s < 2 \pi $. \ 
 $\nu$ is  related  to the $S^5$ spin 
by  
$\displaystyle{J ={ \sql \ov 2 \pi}  \int^{2\pi}_{0}   
d \s\ \del_0 \phi =  \sql \ \nu}$,  i.e. $\nu=\J$. 
For large   $\mu$ it    is related to the $AdS_5$ spin 
$\displaystyle{ S = { \sql \ov 2 \pi}  \int^{2\pi}_{0}  
d \s\ \sinh^2 \r\  \del_0 \theta= \sql\ \S}$ by 
\be\la{spi} 
\mu=\frac{1}{\pi}\ln \S  \ , \ \ \ \ \ \ \ \ \mu \gg 1 \ , \ \
 \ \ \ \ \ \ell= {\nu \ov \mu} =\fix \ . 
\la{lii}
\ee
Rescaling  $\s$ by $\mu \gg 1$ we get 
$\rho = \sigma$ and $\mu$ 
 plays the role of string  length 
 $L= 2 \pi \mu = 2 \ln \S  \gg 1 $
  that scales out of the classical
action and  quantum corrections. 
For $L \to \infty$ the closed  folded string 
becomes effectively a combination of two infinite open 
strings (see also \ci{krtt}).
%N
\foot{For $J=0$ the ends of the string 
that reach the $S^3$ boundary  may be thought of as point  
particles following massless geodesics in $AdS_5$  at $\rho=\infty$
 each carrying  half of the  infinitely large energy  and spin $E= S$, 
while the interior  of the string carries  the extra  energy 
(``anomalous dimension'')
given by string mass, i.e.  the tension times the string length, 
$E-S= { \sql \ov 2\pi}  L = { \sql \ov \pi} \ln \S$.
}

In the scaling limit \rf{ho} the classical energy 
$\displaystyle{ E_0 = { \sql \ov 2 \pi}  \int^{2\pi}_{0}  
  d \s\ \cosh^2 \r\  \del_0 t = \sql\ \E_0}$
  becomes equal to $\E=\S + \k= \S + \sqrt{ \mu^2+\nu^2} $. 
Thus, while   the classical energy 
$\E_0(\S,\J)$ in \rf{ghi}  as a function of two general arguments cannot be written 
in a simple closed form (it is a solution of a system of two parametric 
equations involving elliptic functions \ci{ft1,bfst}), 
in the scaling  limit it simplifies to \ci{bgk}
\be \la{cla}
{\rm SCS:} \ \ \ \ \   \E_0 - \S = \ \mu \sqrt{ 1 + \jj^2}  
= \ { 1 \ov \pi} \ln { \S } \ \sqrt{ 1 +  { \pi^2\J^2 \ov \ln^2 { \S } }   } \ . 
\ee
Since in this limit $\S\gg \J$  and ${S\ov \sql } \gg 1$ 
it is not possible to distinguish between 
$\ln \S $, $\ln (\S/\J)=\ln (S/J)$
or $\ln S$.
%. Moreover, since we are interested only 
%in the leading term in the  large $S$ expansion it is also not possible to distinguish between 
%$\ln \S$ and $\ln S$. 
Therefore, the energy in the ${\rm SCS}$ limit can be also  written as\foot{
Note that there is a  similarity between this  scaling limit of a  folded 
string on $AdS_3 \times S^1$ 
 and the giant magnon
 limit \ci{hm,dor} of a folded string on $R_t \times S^3$.
  As discussed in \cite{mtt}, one can understand 
the latter as the infinite spin limit of a folded $(J_1,J_2)$ 2-spin solution 
on $S^3$ where   one takes $E$ and  $J_2$ to infinity while 
 keeping their difference finite.  Then 
$E-J_2=\sqrt{J_1^2+\lambda k}, $
where $k$ is a constant which depends on the specifics of the initial
 solution. By starting  with the $(S,J)$ solution one can also   consider 
  the limit where $E$ and $J$ are sent to infinity \ci{mtt}; however, a regular 
  scaling limit appears to be the one of \ci{bgk}  where one sends 
   instead  $S$ to infinity while 
  keeping the ratio  $\jj$ in \rf{ho} fixed. The energy then takes again the same 
  universal square root
  form  \rf{kla} (cf. \ci{mtt}).}
 \bea 
{\rm SCS:} \ \ \ \ \    E_0 - S =  { 1 \ov \pi} \ln { S   } \ 
 \sqrt{ \l  +  { \pi^2 J^2 \ov \ln^2 { S  } }   } + ... \ . \la{kla}
 \eea 
  
  \bs 
  
Considering further 
 the SCSS limit \rf{gh}, i.e. expanding in powers of $\jj$   we find
 \be \la{uu}
{\rm SCSS:} \ \ \ \ \ \ \ \ E_0-S &=& \   { \sql  \ov \pi} \ff_0 (\jj)\  \ln { S } + ... \ , \\
\ \ \ \ \ \ \ \ \
 \ff_0 (\jj)&=&\sqrt{ 1 + \jj^2}=  1 + \ha \jj^2 - { 1 \ov 8} \jj^4 + ...
      \ , \la{uuuu} \ee 
or,  equivalently,  
\be
{\rm SCSS:} \ \ \ \ \ \ \  E_0-S = \   { \sql  \ov \pi} \ln { S } \ 
 \big[  1 +   {\pi^2 J^2 \ov 2\l \ln^2 S  } 
 -   {\pi^4 J^4 \ov 8\l^2 \ln^4 { S } } +  ... \big]  + ... \ . \la{ou}
 \ee
The 1-loop  string correction $\E_1$  in \rf{ghi}  was so far 
computed  only in the SCS limit  where it  takes  the form   \ci{ftt}:
\be
\label{oop}
{\rm SCS:} \ \ \ \ \ \ \ \ 
E_1= 
%{ \sql  \ov \pi}  
\ff_1 (\jj) \  \ln { S }\     + ...  \ , \ \ 
\ee  
\bea \la{jl}
 &&\ff_1( \jj ) =   { 1 \ov  \sqrt{ 1 + \jj^2} }
 \bigg\{ \,  \sqrt{ 1 + \jj^2}  -  1
 + 2 ( 1 + \jj^2 ) \ln (1 + \jj^2)  -   \jj^2   \ln \jj^2  
 \no\\ && ~~~~~~~~~~~~~~~~~~~~~~~~~~~~~~~~~~~~~~~~~~~~~~~
  - \ 2( 1 + \ha  \jj^2) \ln \Big[   \sqrt{ 2 +   \jj^2}
    (1 + \sqrt{1 + \jj^2})\Big] \bigg\}   \  . \la{fa}
 \eea
 Expanding  in  small $\jj$ we get from \rf{fa}
 \be
\label{dop}
{\rm SCSS:} \ \ \ \ \ \ \ \
 \ff_1 (\jj) 
 = -{3 \ln 2}  - 2 \jj^2 ( \ln \jj   - { 3 \ov 4}) 
 + \jj^4 ( \ln \jj   - { 3 \ov 8}   \ln 2   - { 1 \ov 16} )
 + O(\jj^6) \ . \ee
% Here the subleading constants  at each order in $\jj^{2n}$
% can be ignored   since $|\ln \jj| \gg 1$ in the SCSS limit. 
 These explicit string-theory results suggest that   the strong-coupling 
 expansion 
 in the scaling limit    may  be organized, following     \ci{am2}, as 
 \bea \la{expa}
 E-S = \  {\sql\ov \pi} \   \ff ( \jj, \l) \  \ln S + ...\ , \\
  \ff = \ff_0(\jj) + { 1 \ov \sql} \ff_1(\jj) + { 1 \ov (\sql)^2} 
  \ff_2(\jj) + ...\ . \la{fef}
  \eea
 In the SCSS limit $ \jj \ll 1$  we may further 
 expand in powers of $\jj$  and organize the expansion as 
 %AT
 \be 
 && \ff ( \jj, \l)
 = f(\l) + \jj^2 \big[ q_0 (\l)  + q_1(\l) \ln \jj  
 + q_2(\l) \ln^2 \jj 
 + ...\big]\no \\
&&  \ \ \ \ \ \ \ \ \ \ \ \  \ \ \ \ \ \ \
 +\  \jj^4 \big[ p_0 (\l)  + p_1(\l) \ln \jj  + ...\big] +  ....\ .  \la{hjk}
 \ee
 Here $f(\l) $  corresponds on the gauge theory side 
 to  the universal scaling function $f$ (expected
 %AAT
  to be the same \cite{be,bes} as the   twist-2 anomalous 
 dimension and thus \ci{kom} the same as the 
 cusp anomalous 
 %AAT
 dimension).\foot{The universality of the minimal  
anomalous dimension of the $(S,J)$ operator at large $S$
 and its relation to
the cusp anomalous dimension was discussed in \ci{bog}.}
  $f$ and 
   the functions $q_n,p_n,...$  
 receive corrections order by 
 order 
 in the inverse tension 
 expansion.
 %  with  $q_1,p_1,...$ starting  with  the 1-loop   string contributions.
 Explicitly,
 \be \la{exl}
 && f(\l) = 1 - {3 \ln 2 \ov \sql}  - {K \ov (\sql)^2}  + 
 ...\ ,\la{feg}
 \\
 &&  q_0(\l)= \ha  +  {3 \ov 2\sql} + ...\ , \  \ \ \ \ \ \  \ 
 \ \ \ \ \ \ \ \  \ \ \ \ \ \ \ \ \,
 q_1(\l)= - {2 \ov \sql}      + ...\ , \la{exc}  \\
 && 
 p_0(\l) = - { 1 \ov 8} -   {1 \ov 8\sql}( 3   \ln 2   + { 1 \ov 12} )   ...\ , \ \ \ \ \ \ \ \ \ 
 p_1(\l)= {1 \ov \sql}+...\ . 
 \ee
 ${\rm K}$ in $f(\lambda)$  
 is the Catalan's constant, i.e. 
 we included   2-loop string-theory result   \ci{rtt,rt}  
  %AAT
  (which matches the corresponding  strong-coupling expansion 
 coefficient found from the BES equation \ci{bkk}).

 \bs

 The formal expansion in \rf{hjk} is to be understood 
 in the sense that  higher powers of $\ln \jj$ are suppressed
 compared to lower  ones at each given order of the  strong   coupling expansion
 in which we {\it  first}
  take $\l\gg 1$  and {\it then}  take $\jj$ to be small, 
  i.e.  in  the semiclassical  limit in string theory  we are considering 
  we should really 
 to use \rf{fef}  where   each of the loop corrections 
  $\ff_n(\jj)$  is then expanded in 
 $\jj$, i.e. \rf{uu} and \rf{dop}. 
  
 %It is  not  clear {\it a priori}  whether  such  representation \rf{hjk} 
 % where one  expands in $\jj$ 
 %while keeping  the coefficients  of different structures as  functions of $\l$ 
 % should make sense in general:
 %in  the semiclassical  limit in string theory  it is natural 
 %to use \rf{fef}  where   each of the loop corrections  $\ff_n(\jj)$  is then expanded in 
 %$\jj$, i.e. \rf{uu} and \rf{dop}. 
 %However,   to compare to gauge theory  where $S$ and $J$ are kept fixed when $\l$ is increased 
 %it seems  that it is desirable to re-expand as in \rf{hjk}.
 
 %N
 
 \bs 
 As was  argued in \ci{am2}, in the $O(6)$ model limit \rf{amm}   one
 should be able to determine the coefficients  of the two leading  logaritmic terms in 
 in the  large $\lambda $ expansion of $\ell^2$ term of 
 \rf{hjk}. The  input is  the expression  of free energy of the $O(6)$ model \ci{has,EH} 
 and the string-theory  value of the 1-loop coefficient in $q_0$ in  \rf{exc}
 that determines the value of $\kk$ in \rf{scm}. The resulting prediction is \ci{am2}
 \be \la{ggff}
 {\rm AM}: &&\ \ \ \  \  \ \ \ \ \ \ \ 
 \ff ( \jj, \l) = \ \ell^2 \sum^\infty_{n=1} {1 \ov (\sql)^{n-1}} \ 
  \Big( c_n \ln^n \ell + d_n \ln^{n-1} \ell
 + ... \Big)  \ ,  \\
   \la{pred}
&& c_1 = -2 \ , \ \ \   d_1= { 3\ov 2} \ , \ \ \ \ \ \ \ 
  c_2 = 8  \ , \ \ \   d_2= - { 6} \ , \ \ \ \ \ \ \ 
   c_3 = -32  \ , \ \ \   d_3= 12 \ ,  ...   \ee 
 Here $c_1,d_1$ follow from \rf{fa} and $c_n,\ d_n$ with $n>1$ are the values 
 determined \ci{am2} by the $O(6)$ model in the limit \rf{amm}.\foot{The values of $c_n$ 
 are essentially determined by the value of the 1-loop coefficient $c_1$, while 
 the values of $d_n$ are sensitive to the value of the ``string'' 1-loop 
 constant $d_1$  or  $\kk$ in \rf{scm}.}
  %according to \ci{am2}.

 \bs

 The presence of $\ln^n \jj$ terms in \rf{hjk}
  appears to be an artifact of inverse 
 tension or loop expansion on the string side:  to compare to gauge theory one would 
 need  to resum the logarithmic terms (see  also section 5). 
 On the 
 %An important point is that the expansion in small $\jj$  has a chance to match the 
 %strong coupling expansion on 
  gauge side  the limit is taken 
 in a different way:
 by  first considering    $\ln S \gg J$  at small $\l$ and then continuing 
  to large $\l$. 
 Since    $\jj = { \pi J \ov \sql \ln S}\equiv {j\ov \sql} $ 
 is naturally  small also in {\it that} limit,  
  the two  limits may  actually commute.

Let us add that the  1-loop string expression \rf{fa} was 
reproduced  \ci{krj} from the  ``string'' form of the Bethe ansatz 
  (with the   phase \ci{afs}  in the $S$ matrix expanded  at strong coupling \ci{bhl}
   with the 1-loop term in it 
   determined \ci{bt,hl}  from some other 1-loop string results).
 For work towards the  determination of the weak-coupling 
 gauge theory predictions in  the scaling   limit for non-zero $\jj$ 
  see  \ci{bgk,belit} and especially \ci{staud}.

\bs\bs
 
 Our aim here   will be to compute $\ff_2$  in \rf{fef} in the SCSS limit,
   i.e. to
 determine   how the expansion \rf{hjk} is modified 
 by the {\it 2-loop}  string corrections. 
 We shall  find that 
 the coefficient of the $\jj^2$ term   in \rf{hjk}  receives 
 $\ln^2 \jj$ corrections 
 % was  suggested  in 
 %\ci{am2} 
  and the coefficients of the 2-loop 
 $\ln^2 \jj$ and $\ln \jj$ terms are exactly the same as 
  the  values of $c_2$ and $d_2$ in \rf{pred} predicted  in \ci{am2}. 
 % by  
% analogy with  the $O(6)$ model 
%  origin of the 1-loop $\jj^2 \ln \jj$ term in \rf{dop}
 % coming from the  light   $S^5$   modes. 
 
\bs 

The rest of this paper is organized as follows. 
In section 2 we shall explain the general procedure for 
 computing quantum  corrections 
to the space-time 
energy of the string  with fixed values of the spins. 
In section 3 we shall discuss
 the form of the classical solution and the  \adss action that
  we will be using for the 2-loop 
computation of the energy in the scaling limit. 
Details of  the computation of the 
 1-loop and the 2-loop  quantum contributions to the 
world-sheet effective action will be discussed  in section 4.  

In section 5 we shall 
present our  final result for the 2-loop term in \rf{fef}
and show its  agreement  with  the prediction made in \ci{am2}. 
%N
In section 6 we 
shall  suggest that it should be possible to resum all
the $\jj^2 \ln^n \jj$  terms  so  that they  should 
disappear in the weak-coupling gauge theory limit. 
 %Few concluding 
%remarks will be  made in section 6.  
Appendix A will contain the  results for 
some relevant 2-loop integrals.
  In Appendix B
we will consider  a model computation of the 2-loop  correction to 
the effective action  of the $S^2$  sigma model.

%%%%%%%%%%%%%%%%%%%%%%%%%%%%%%%%%%%%%%%%%%%%%%%%%%%%%%%%%%%%%%%%%%%%%%%%%%%%%%%%%%%%%%

\section{Structure of computation of quantum corrections to string energy}
%%%%%%%%%%%%%%%%%%%%%%%%%%%%%%%%%%%%%%%%%%%%%%%%%%%%%%%%%%%%%%%%%%%%%%%%%%%%%%%%%%%5

As explained above, 
 we would like   to compute the 2-loop string correction to the energy $E$ of the 
spinning string as a function of the spins $S$ and $J$ in  the scaling limit \rf{ho}. 
The 2-loop order in the multi-spin case is the first  time   when 
 we  face an important conceptual subtlety that was not addressed 
in  an earlier work. 
  Similarly to the energy,  the spins  may 
  receive quantum corrections from their  classical values 
 given  
 %(up to the string tension  $\sql \ov 2\pi  $  factor)
  by the semiclassical ``frequency'' 
 parameters.  In general, to  be able to compare to  gauge theory 
 we need to  find the energy $E$ as a function of the exact values of the spins, i.e. 
 $E$  should be computed with these  values fixed. 
 In fact, we should treat the energy  and the spins  (i.e.  all of the 
 $SO(2,4) \times SO(6)$
 charges)   on an equal  footing,  relating them at the end via the 
 quantum conformal gauge constraint.

 It was argued  in \ci{ftt,rtt,rt} that in the scaling limit, i.e.  in the limit 
 of large  string length, 
  the quantum string correction  to the space-time energy  (i.e. the global $AdS_5$
   energy) can be computed 
 from  the string partition function. 
 We will further justify this  below; a key
 ingredient of the argument is  that the volume of the string world sheet 
 is large and factorizes. 
 
 In the quantum theory, the symmetry charges are found as expectation values of 
 integrals of the corresponding N\"oether currents.
 In the scaling limit which  we are considering  here (keeping only the leading 
 power of 
 the effective string length or $\mu= {1\ov \pi} \ln \S$)  we need to take 
 into account only  corrections to
 the second  ``small'' spin $J$.
 %\foot{
 This is  so because the perturbative
 corrections to the expectation value of any operator 
 %(in particular,  
 %the spin $S$ operator) 
 can
 %not 
 depend 
 %directly 
 %on $S$ but  
 only on the parameters of the classical solution like $\kappa$ or $\mu$ 
 which can 
 grow with $S$  at
 most as $\ln S$. Consequently, the corrections to the spin $S$ operator (and
 thus  their contributions  to the energy)
  are suppressed by the inverse powers of
 $S$.
 %}
 
 The main  observation is  that
 %, from the standpoint of the
 %world sheet theory, 
 the space-time energy  as well as the   spins  
 in $AdS_5$ or $S^5$ 
 are  conserved charges of the world-sheet theory:  they  correspond to the 
   generators of the $PSU(2,2|4)$ global symmetry group of the string sigma model. 
%
% Thus they should be fixed   while  computing the  string partition function.
%
 Thus, they should be treated on an equal footing in a conformal-gauge 
 world sheet calculation. The global conserved charges are related by 
 the Virasoro constraint, i.e. by  the requirement that the 2d world-sheet  energy vanishes.

 %AT
Let us 
%%add a 
comment on the  implementation of the Virasoro constraint in a 
semiclassical expansion. Since one is to expand around a solution of the classical equations of motion, one should impose the classical Virasoro constraint. Both the 
charges and  the world sheet energy receive quantum corrections order by order in 
sigma model
perturbation theory.
 They are naturally expressed in terms of the parameters of the classical solution. The quantum Virasoro constraint requires the vanishing 
 of the world sheet energy. This constraint therefore imposes an additional 
 relation between the parameters of the classical solution which can be 
 satisfied only if the quantum corrections to the global symmetry charges are correlated
 in a certain way. 
It is this relation that we are going to expose and exploit below.  
 
 The implementation of the fixed-charge constraint is a well-known problem in 
 statistical mechanics  leading  to the notion of generalized ensemble. 
 Below we shall first review some relevant
  general points and then turn to our specific string 
 theory sigma model case. 
 
%%%%%%%%%%%%%%%%%%

\subsection{Partition function with  fixed charges \label{ZQ}}
 
In general, one thinks of the partition function as a sum over states. It is
 typically hard to sum only over the states of some definite fixed charge -- 
especially when one does not know all the states of the theory. This difficulty led 
to the concept of generalized statistical ensemble. The main idea is that, 
instead of using charges to label states, one interprets the charge as an 
average quantity which can then be set to any desired value.\footnote{Clearly, this 
standpoint is particularly appropriate if one expects that
the  symmetry generators receive quantum corrections.}

Let us review the definition and construction of the corresponding partition function.
We consider a system whose states have some energy $E$ and carry some charge 
$Q$. By definition, the partition function is simply the normalization factor 
of the probability
that the system has  energy $E$ and  charge $Q$. To find this probability
 we start with the fact that the number of states for the system in contact
with a reservoir of energy 
and charge at fixed {\it  total} energy $E_\rT$ and 
{\it total }  charge $Q_\rT$ is
\be
\Omega_\rT(E_\rT,Q_\rT,\dots)&=&\int \frac{dE}{\Delta}\sum_{Q}\Omega(E,Q)\ 
\Omega_R(E_\rT-E,Q_\rT-Q)
\cr
&=&\int \frac{dE}{\Delta}\sum_Q \ \Omega(E,Q)\ 
\exp\left[
%\textstyle{
%\frac{1}{k_{_B}}}
S_R(E_\rT-E,Q_\rT-Q)\right] \ . \la{zx}
\ee
%R
Here $\Omega$ denotes the number of states of the system, $\Omega_R$ the number of states
 of the reservoir, $S_R$ is its entropy
and $\Delta$ is a coarse graining parameter which is arbitrary, apart from the 
fact that it should not scale with the volume of the system.  
It follows then that 
the probability for the system to have the energy $E$ and charge $Q$ is
\be
P(E,Q)\propto \Omega(E,Q)\ \exp\left[
%\textstyle{
%\frac{1}{k_{_B}}}
S_R(E_\rT-E,Q_\rT-Q)\right]\ .  
\ee
We may then expand the entropy of the reservoir $S_R(E_\rT-E,Q_\rT-Q)$, using 
the fact that its energy and charge (and therefore the total energy
$E_\rT$ and total charge $Q_\rT$) are by definition much larger than the energy 
and charge of the system:
\be
S_R(E_\rT-E,Q_\rT-Q)&=&S_R(E_\rT,Q_\rT)+\frac{\partial S_R(E_\rT,Q_\rT)}
{\partial E_\rT}(-E)+
\frac{\partial S_R(E_\rT,Q_\rT)}{\partial Q_\rT}(-Q)+ ...
\cr
&\approx &S_R(E_\rT,Q_\rT) -\beta E + \beta  h Q \ , 
\la{ok}
\ee
where $\beta$ is  the inverse  temperature and
 $h$ is the variable canonically conjugate to the charge
 (e.g.,  if $Q$ is an electric charge then $h$ is an electric potential).
  We therefore find that
the probability for the system to have energy $E$ and charge $Q$ is given 
(up to overall factors independent of $E$ and $Q$) by 
\be \la{ujk}
P(E,Q)\ \propto\  \Omega(E,Q)\ e^{-\beta (E-h\, Q)}~~.
\ee
Assuming that $E$ and $Q$ are the eigenvalues of 
the {\it commuting} operators ${\widehat H}$ and ${\widehat Q}$, the 
partition function is
\be
Z( h)=\Tr[e^{-\beta ({\widehat H}-{h}\,{\widehat Q})}] \ , \la{yyy}
\ee
where the trace is taken over all states.\footnote{The presence of $\beta$ 
in front of both ${\widehat H}$ and $h\,{\widehat Q}$ implies that 
${\widehat H}-h\,{\widehat Q}$ is the evolution operator in this ensemble.
We suppress the obvious argument $\beta$ 
in $Z$ and related quantities.}

 This implies,  on general grounds, 
   that the logarithm of the partition function
is related to $\langle {\widehat H} \rangle $ and $\langle {\widehat Q} \rangle $ as
\be
 -\frac{1}{\beta} \ln Z(h)=\langle {\widehat H} \rangle 
                          - h \langle{\widehat Q}\rangle
 \equiv \Sigma(h)~,
\label{GD}
\ee
which defines the generalized thermodynamic potential ${\Sigma(h)}$ 
(which may also be loosely called free energy).

Then, the average value of the energy $\langle {\widehat H}\rangle$ over the 
states with fixed charge $Q$ is the Legendre transform of $\Sigma$ with respect to $h$. 
Namely, we are first to compute the partition function with the modified
Hamiltonian
\be\la{uuu}
 {\widehat {\widetilde H}}={\widehat H}-{h}\,{\widehat Q}\ , 
\ee
from which we are to find the  value of the charge averaged  over all states
\be\la{kkl}
\langle \widehat Q\rangle = \frac{1}{\beta}\frac{\partial \ln Z(h)}{\partial h}~~.
\ee
Setting this to the desired value 
\be 
\langle {\widehat Q}\rangle = Q\  , 
\ee
we find $h=h(Q)$. Then the  energy averaged over the states with fixed charge
is found by evaluating
\be
 \langle {\widehat H}\rangle=-{1 \ov \beta} \ln Z(h(Q)) + h(Q)\,Q  \ . \la{jip}
\ee
In a semiclassical expansion $\langle {\widehat H}\rangle$ comes out as a series 
of quantum corrections to the classical energy.

This construction is completely general  under the assumption that the volume of the
 system is large (infinite) and that the interactions are local.\footnote{Corrections
  to the equation \rf{GD} are suppressed by inverse powers of the volume. Moreover,
  the root-mean-squared fluctuations of the energy and the charge around their average
  values are also suppressed by inverse powers of the spatial volume. Finally, 
  one may also argue that under these conditions $\langle {\widehat H}\rangle$ and 
  $\langle {\widehat Q}\rangle$ are the most probable values of the energy and the 
  charge.} 
It can of course be generalized to include two or more    different charges. 

\

In the following we  will identify the ``undeformed'' Hamiltonian ${\widehat H}$ 
with the world-sheet Hamiltonian, one charge with the difference between 
the space-time energy $E_{\st}$  and the $AdS_5$ spin $S$ 
and a second charge with the angular momentum  $J$ 
on $S^5$. The reason for using $ E_{\st}-S$ as a charge instead of 
introducing separately 
$E_{\st}$ and $S$ is 
%due to the fact 
that in the present case this difference is
 parametrically smaller that either $E_{\st}$ or $S$. 
 %AT
 %We will 
 %,  however,  that we will 
%encounter some  interesting subtleties related to the fact that 
%we are dealing with  a world sheet theory.

\subsection{Partition function of a world-sheet theory with fixed charges \label{WSfq}}

The discussion of   section \ref{ZQ} can be easily  translated to  
field-theory language and applied to the  semiclassical expansion 
of a two-dimensional string sigma model.

As explained above, we are to consider the world-sheet 
Hamiltonian modified by the 
addition of the charge operators.
 %
 %R
 %
 Anticipating the relation to  the equation \rf{solc}, 
 we shall denote by $-\kappa$ the  ``chemical potential''  $ h$  conjugate 
 to the difference between the space-time energy $E$ and the $AdS_5$ spin $S$
  and by $\nu$  -- the  ``chemical potential'' 
  conjugate to the $S^5$ spin  $J$: \footnote{The sign difference between $h_{_{E-S}}=-\kappa$
  and $h_{_J}= \nu$ 
   has to do with the fact that $\kappa$ is in a sense 
   ``time-like'' -- much like the entropy: the variable conjugate to the entropy
    is $\b$ while the one conjugate to the  charges is  $- \b h $.}
\be\la{kou}
{\widetilde H}_{\ws}=H_{\ws}+\ \kappa\,(E_{\st}-S)-\nu\,J  \ . 
\ee
The partition function with this modified Hamiltonian may be computed 
by transforming  it first in a standard way into a  path
 integral   with  euclidean  action.\foot{For a 
 %AAT
  discussion of  
 a formally similar definition  of 
  free energy in  two-dimensional  sigma models see  \ci{pw,has,fat,for}.
  The present string  case will, however, be conceptually different in that we will have the 
  Virasoro constraint.}
 
 To this end we note that $E_{\st}$, $S$ 
and $J$ are  momenta conjugate to world sheet fields 
 -- the global
 time direction $t$, an isometric angle $\theta$ in $AdS_5$ and an 
 isometric angle $\phi$ on $S^5$
 (see  \rf{solc}). It is then 
easy to see that  the Lagrangian associated to the 
modified Hamiltonian ${\widetilde H}_{\ws}$ is  obtained from the one 
associated to $H_{\ws}$ simply by shifting the time derivatives of the relevant fields 
by constants $\kappa$ and $\nu$:
\be
{\dot t}\mapsto {\dot t} +\kappa
~,~~~~~~~~
{\dot \theta}\mapsto {\dot \theta} +\kappa
~,~~~~~~~~
{\dot \phi}\mapsto {\dot \phi} +\nu \ . 
\label{repl}
\ee
An  expansion around any classical solution of this modified Lagrangian
is then  found by  including an additional background 
 term  in each of the fields $t,\theta$ and $\phi$ 
\be
{ t}=\kappa\tau  +  {\td t}(\sigma, \tau) \ , 
~~~~
~~~~
{\theta}=\kappa\tau + {\td \theta} (\sigma, \tau)\ , 
~~~~
~~~~
{ \phi}= \nu\tau  + {\td \phi}(\sigma, \tau) \ .  \la{iiy}
\ee
One
% is then to add   if needed 
may want to include 
 also  the profiles of other fields 
(such as $\rho$ in  \rf{solc}).

Since we want to interpret the resulting Lagrangian as that  of a  world-sheet theory, we need 
to identify the corresponding  Virasoro constraints (we shall use the conformal gauge). 
To this end we note that on a curved world sheet
the momentum conjugate 
to a field contains a factor of the inverse metric. This implies that the 
replacement \rf{repl} is sufficient also in the presence of a nontrivial world 
sheet metric  and thus  the Virasoro
 constraint for  the modified Lagrangian follows from that for 
 the original one by the same replacement \rf{repl}. 
The Virasoro condition 
 then relates $\kappa$ and $\nu$ (cf. \rf{Vir}) 
\be \la{viri}
\kappa=\kappa(\nu)~.
\ee 
Consequently, the partition function is a function of only $\nu$; 
therefore,
it is impossible to ``measure''  separately $\langle E_{\st}-S \rangle$ and 
$\langle J\rangle $. Instead 
of $\Sigma(\kappa,\nu)$ in \rf{GD} we  get
$\Sigma(\nu)\equiv\Sigma(\kappa(\nu),\nu)$ and 
 its derivative  is a combination of 
  $\langle E_{\st}-S \rangle$ and  $\langle J\rangle $
\be
\frac{d\Sigma(\nu)}{d\nu}&=&
\frac{\partial \Sigma(\kappa,\nu)}{\partial \kappa}\Big|_{\kappa=\kappa(\nu)}\,
\frac{d\kappa(\nu)}{d\nu} + 
\frac{\partial \Sigma(\kappa,\nu)}{\partial \nu}\Big|_{\kappa=\kappa(\nu)} 
\cr
&=&
%\tT 
 \frac{d\kappa(\nu)}{d\nu}
\langle E_{\st}-S\rangle  - \langle J\rangle\ . 
\label{rel1}
\ee
A  second relation between these quantities is found by recalling that 
the expression for  the generalized potential in terms of the average values of 
charges is 
\be
\Sigma(\nu) = 
%\tT 
\langle H_{\ws}\rangle + 
 \kappa(\nu) \langle E_{\st}- S\rangle 
-  \nu \langle J\rangle \ , 
\label{GDJ}
\ee
and imposing the  quantum Virasoro constraint
\be\la{vvv}
\langle H_{\ws}\rangle=0 \ . 
\ee
%This  gives another  relation  between
% $\langle E- S\rangle $, $\langle J\rangle$ 
 $\Sigma(\nu)$  
 is proportional to the  world-sheet effective action $\Gamma(\nu)$
 \be \Gamma(\nu) \equiv -\ln Z(\nu)=  \beta \Sigma(\nu) \ . \la{efg} \ee 
 Here  we will be interested  in the 
zero temperature limit  when  $\beta\equiv \tT  \to \infty$ plays the role of the 
 length of the world-sheet  time interval.

Combining (\ref{rel1}) and (\ref{GDJ}) and setting the average values of the 
charges 
% $E-S$ and   $J$ 
 to the desired values, 
 \be
 \langle E-S\rangle=E-S\ , \ \ \ \ \ \ \ 
 \langle J\rangle=J\ , \ee
   we find that
\be
 E_{\st} -S
&=&\frac{1}{\tT}\left(\nu\frac{d\kappa(\nu)}{d\nu}-\kappa(\nu)\right)^{-1}
\left(\Gamma(\nu)-\nu\frac{d\Gamma(\nu)}{d\nu}\right) \ , 
\la{ej} \\
J&=&\frac{1}{\tT}\left(\nu\frac{d\kappa(\nu)}{d\nu}-\kappa(\nu)\right)^{-1}
\left(\Gamma(\nu)\frac{d\kappa(\nu)}{d\nu}-\kappa(\nu)\frac{d\Gamma(\nu)}{d\nu}\right) \  . 
\la{pq}
\ee
%where as before $T$ is the length of the time direction. 

Before  proceeding let us point out that
 the discussion in this section has a trivial  generalization 
to the multi-spin cases  where  instead of one  parameter $\nu$ we have 
several of them;  then 
the Virasoro  condition implies  that 
\be
\kappa=\kappa(\nu_1,\dots,\nu_n) \ ,  
\ee
 and the equation \rf{rel1} generalizes to a system of equations:
\be
\frac{d\Sigma(\nu_1,\dots ,\nu_n)}{d\nu_i}
&=& \frac{\partial \kappa(\nu_1,\dots, \nu_n)}{\partial \nu_i}
\langle E_{\st}- S\rangle -\langle J_i\rangle \ , \ \ \ \ \ \  ~~~~i=1,\dots, n~~.
\label{relgen}
\ee
Eq. \rf{GD} combined with the quantum Virasoro constraint becomes
\be
\Sigma(\nu_1,\dots,\nu_n) =  
 \kappa(\nu_1,\dots \nu_n)\langle E - S\rangle
-  \sum_{i=1}^n \nu_i \langle J_i\rangle  \ . 
\label{GDgen}
\ee
For independent charges $J_i$, this system of $(n+1)$ equations for the same 
number of unknowns is nondegenerate.

\subsection{Sigma model loop expansion \label{ZQVir}}

Let us now specify the discussion of  the previous section to our 
particular solution \rf{solc}  with $\rho= \mu\ \sigma$,\ \ $\mu \to \infty$.  
To make manifest the fact that it
applies without modification we will rescale the world sheet coordinates  
$\tau$ and $\sigma $ by $\mu$. The homogeneity of the solution, reflected in 
 the fact that  the coefficients in the fluctuation Lagrangian are 
constant  \ci{ftt,rt}, guarantees 
that $\mu$ can be rescaled out in 
 the classical action with   the parameters $\kappa$ 
and $\nu$  replaced by
\be\la{pas}
\hkappa = \frac{\kappa}{\mu}~, \ \ \ \ \ \ \ \ \ \ ~~~~~~~~~~\lnu=\frac{\nu}{\mu} \ .
\ee
Then  the quantum effective action is proportional to the world-sheet 
volume 
\be\la{gas}
&& \Gamma = 
%{ \sql \ov 2 \pi} 
{ \sql \ov 2 \pi}  \ V_2\   {\cal F}(\lnu) \ ,  \\ \la{vol}
&&V_2(\mu) = 2\pi \mu\; \tT~,\ \ \ \ \ \ \ \ \tT= \mu \,{\bar \tT} \ , 
\ \ \ \ \  \ \ \ \ 
 \mu = { 1 \ov \pi} \ln \S \approx { 1 \ov \pi} \ln {S} \gg 1  \
.
%~~~~T=\mu {\bar T}
\ee
${\cal F}(\lnu)$  has  a standard expansion in inverse powers of $\sql $ with a 
constant leading term.

Using \rf{Vir} after the the rescaling of world sheet coordinates by $\mu$, i.e. 
 \be\la{yt}
\hkappa(\lnu)=\sqrt{1+\lnu{}^2}\   \ee
 and evaluating \rf{ej} and \rf{pq} it is 
easy to find $E-S$ and $J$ in terms of ${\cal F}(\lnu)$ and its first derivative:
\be
E-S&=& \ \cM \sqrt{1+\lnu{}^2}\ 
\bigg[ {\cal F}(\lnu)-\lnu\frac{d{\cal F}(\lnu)}{d\lnu}\bigg] \ , 
\la{q} \\
J 
&=&  \ \cM \bigg[ \lnu{\cal F}(\lnu)-
(1+\lnu{}^2) \frac{d{\cal F}(\lnu)}{d\lnu}\bigg] \ , \la{o}
\ee
where $\cM = {\sql \ov 2 \pi} L$ is the ``string mass'' (tension 
times length),\foot{Recall that,  
as discussed in section \ref{intro}, 
the scaling  limit $\sql \ln S \gg J , \ \ S \gg \sql,\  \l \gg 1, $
 makes $\ln \S$, $\ln S$ and $\ln S/J$ indistinguishable.}
\be \la{my}
\cM \equiv    { \sql \ov 2 \pi} { V_2 \ov \tT}    
= {\sql  \ov \pi} \ln S 
\ . \ee
Expanding in the inverse string tension  we get 
\be
\la{uo}
{\cal F}(\lnu)={\cal F}_0+\sum_{n=1}^\infty { 1 \ov  (\sql)^{n} }
 {\cal F}_n(\lnu)\ .
\ee
As we shall see below, 
\be \la{hl}   {\cal F}_0 =1   \ . \ee
Introducing the notation for the analog of $\ell$ in \rf{ho} or \rf{fi} 
(and using \rf{hl}) 
\be \la{nota}
\jjj=\frac{J}{ {\cal F}_0\ \cM  
}=
%\frac{1}{{\cal F}_0}\,\frac{\pi J}{\sqrt{\lambda}\,\ln(S/J)} = 
\frac{\pi J}{\sqrt{\lambda}\,\ln{S } }  \ , 
\ee
we find that the first few orders in the inverse  tension 
 expansion of the energy have the same structure as in \rf{expa}:
\be
E-S &=&\frac{\sqrt{\lambda}}{\pi}\ln S\ \ \Big[
{\tE}_0+\frac{1}{\sql }{\tE}_1+\frac{1}{(\sql)^2}{\tE}_2+\dots\Big]\ , \la{oo} \\
{\tE}_0&=&{\cal F}_0\sqrt{1+\jjj^2}\vphantom{\Big|} \ , \la{j} \\
{\tE}_1&=&\frac{{\cal F}_1(\jjj)}{\sqrt{1+\jjj^2}} \ ,  \la{gg} \\ 
{\tE}_2&=&\frac{1}{\sqrt{1+\jjj^2}}\left[{\cal F}_2(\jjj)
+\frac{1}{2 {\cal F}_0}\left(\frac{\jjj}{\sqrt{1+\jjj^2}}\,{\cal F}_1(\jjj)
-\sqrt{1+\jjj^2}\,\frac{d {\cal F}_1(\jjj)}{d\jjj}\right)^2\right] \ . \la{pp}
\ee
It is also straightforward to find the expression  for  $\lnu$ in terms of $\jjj$:
\be
\lnu&=&\lnu{}^{(0)}+\frac{1}{\sql} \lnu{}^{(1)}+\frac{1}{(\sql)^2}\lnu{}^{(2)}+\dots\ , \la{pyy}
\\
\lnu{}^{(0)}&=&\jjj\ , 
\vphantom{\Big|}
\la{vnm}\\
\lnu{}^{(1)}&=&\frac{1}{{\cal F}_0}\left[(1+\jjj^2)\frac{d {\cal F}_1(\jjj)}{d\jjj}
-\jjj\,{\cal F}_1(\jjj)\right]  \ , \la{reno}\\
\lnu{}^{(2)}&=&\frac{1}{{\cal F}_0}\left[(1+\jjj^2)\frac{d {\cal F}_2(\jjj)}{d\jjj}
-\jjj\,{\cal F}_2(\jjj)\right] 
+\lnu{}^{(1)}\; \frac{d \lnu{}^{(1)}}{d\jjj}  \ .
\label{h_exp}
\ee
The fact that to the  leading order we have $\lnu=\jjj$
 implies that the SCSS expansion 
\rf{ou}, \rf{hjk} is equivalent to the  small $\lnu$ expansion of the effective 
action.

We are thus to compute the effective action $\Gamma(\lnu)$ or, equivalently, 
the function ${\cal F}(\lnu)$  and then expand it  in small  $\lnu$.

%%%%%%%%%%%%%%%%%%%%%%%%%%%%%%%%%%%%%%%%%%%%%%%%%%%%%%%%%%%%%%%%%%%%%%%%%%%
\section{Classical solution  and fluctuation action}

The calculation of the
 two-loop world sheet partition function in the
  presence of the {\it ``chemical potentials''} $\hkappa$ and $\lnu$
%
%this calculation 
%
is most easily done using the 
 T-dual  version of the \adss string action in the 
  Poincar\'e patch of $AdS_5$.
  %\foot{The use of the formal T-duality (2d duality)
 % transformation of the \adss superstring action \ci{KT}
  % is justified in the scaling limit 
  %for  the quantum fluctuation 
 % Lagrangian defining the quantum corrections 
  %to the string partition  function.}

  Let us  first discuss
  this map for the solution \rf{solc}, following \cite{rt} and \cite{krtt}, 
  and then proceed to construct the constant-coefficient action for 
  the fluctuations around this solution.

\subsection{Solution   and \adss  superstring action in Poincar\'e coordinates}

To transform  the Minkowski-signature folded closed string solution in global coordinates 
into the Poincar\'e-patch Euclidean-signature 
solution 
let us use  the embedding coordinates and  perform a discrete 
$SO(2,4)$ transformation as well as an analytic continuation. 

In the embedding coordinates, the $AdS_5$ part of the solution \rf{solc} is:
\be
&&
X_0=\cosh\mu\sigma \ \cos\kappa\tau
~, ~~~~~~~
X_5=\cosh\mu\sigma \ \sin\kappa\tau\ , \cr
&&
X_1=\sinh\mu\sigma \ \cos\kappa\tau
~, ~~~~~~~
X_2=\sinh\mu\sigma \ \sin\kappa\tau\ . 
\ee
Analytically continuing to the Euclidean world sheet time, 
$\tau=-i\tau'$,
as well as interchanging a space-like and a time-like target space coordinates,
$X_2=i X_5'$, \  $X_5=i X_2'$ (which is a discrete $SO(2,4)$ transformation) 
leads to
\be
&&
X_0=\cosh\mu\sigma \ \cosh\kappa\tau'
~, ~~~~~~~
X'_2=\cosh\mu\sigma \ \sinh\kappa\tau'\ , \cr
&&
X_1=\sinh\mu\sigma \ \cosh\kappa\tau'
~, ~~~~~~~
X'_5=\sinh\mu\sigma \ \sinh\kappa\tau'~.
\ee
Further discrete $SO(2,4)$ transformations in the planes $(0,5)$ and $(1,2)$ 
put this this solution into the form
\be
&&
X_0=\frac{1}{\sqrt{2}}\cosh(\mu\sigma+\kappa\tau')
~,~~~~~~~
X_5=\frac{1}{\sqrt{2}}\cosh(\mu\sigma-\kappa\tau')\ , \cr
&&
X_1=\frac{1}{\sqrt{2}}\sinh(\mu\sigma+\kappa\tau')
~,~~~~~~~
X_2=\frac{1}{\sqrt{2}}\sinh(\mu\sigma-\kappa\tau')~.
\ee
%AT
Finally, interpreting this as a  background in the Poincar\'e patch 
 with the metric 
 $$ds^2=z^{-2}(dz^2+dx^mdx_m)$$
   yields
\be
z=\sqrt{2}\,e^{-\kappa \tau'+\mu\sigma}
~,   \ \ \   ~~~~~~~
x^0=\frac{z}{\sqrt{2}}\,\cosh(\kappa \tau'+\mu\sigma)
~, \ \ \ ~~~~~~~
x^1=\frac{z}{\sqrt{2}}\,\sinh(\kappa \tau'+\mu\sigma) \ . 
\label{solpp}
\ee
 The $S^5$ coordinates are affected only by 
 the analytic continuation to the 
 Euclidean time: the resulting profile for the $S^5$ 
field $\phi$ is
\be\la{iy}
\phi=i\nu\tau'~.
\ee
In the following we will omit the prime,  denoting  the Euclidean 
world sheet time coordinate simply by $\tau$.

Throughout 
the above  chain of transformations we have carefully kept track of the parameters 
$\kappa$ and $\nu$. While from the standpoint of the resulting 
solution their interpretation as chemical potentials for certain global 
charges is obscured, the fact that to find \rf{solpp} we used only symmetries of 
the string action  guarantees that their meaning 
is unchanged.

\
%AT

Eqs. \rf{solpp}, \rf{iy}  define  the solution we will be using from now on. 
To get a  simple {\it quadratic} form of the resulting fermionic part of the fluctuation 
action 
we can view it, following  \cite{rt},   as corresponding to the 
 T-dual form of the \adss superstring action 
 %AAT
  \cite{KT}:\foot{The use of this 
 action, T-dual to the \adss  superstring action in a particular $\kappa$-symmetry gauge, 
 here is  a  technical trick to simplify the computation of the fermionic
 contributions (see also  \ci{rt}). The T-duality  which is a formal 2d duality transformation 
  at the level of path integral is legitimate to use  in the computation of 
  quantum corrections near  a classical solution. 
  %AAT
  This is true 
   provided the solution
   is ``covariant'' under this
  transformation  which is the case here in the scaling limit $\mu \gg 1$ when 
  $\mu$ and $\kappa$ (and  $\tau$ and $\sigma$) enter on an equal footing (see also \ci{rt}).}
\be I&=&I_B+I_F \ , \ \ \ \ \ \ \ 
{I}_B=\frac{\sqrt{\lambda}}{4\pi}\int d^2\sigma\
{\cal L}_B=\frac{\sqrt{\lambda}}{4\pi}\int d^2\sigma\
\frac{1}{z^2}\big(dx^mdx_m+dz^sdz_s\big) \ , 
\la{jjk} \\
{I}_F&=&\frac{\sqrt{\lambda}}{4\pi}\int d^2\sigma\
{\cal L}_F=\frac{\sqrt{\lambda}}{4\pi}\int d^2\sigma\
4\ \epsilon^{ab} \ {\bar\theta} \left( {\partial_a x^m  }  \Gamma_m \ 
+ \   {\partial_az}^s \Gamma_s   \right)  \partial_b\theta \ . 
 \vphantom{{}^|}
\label{action}
\ee
The coordinates $z^s$ are flat coordinates on $\IR^6$; the coset parametrization 
used in  \cite{KT} relates them to a particular
choice of  coordinates on $S^5$ as
($i=4,5,6,8,9$)
\be
&&z^i\equiv\ z\ {\hat z}^i=\
z\ \frac{{\hat y}^i}{1+\frac{1}{4}{\hat y}^2}~, \ \ \ \ \ \   \ \ \ \ \ ~~
~~~~~~~~
z^7\equiv z\, {\hat z}^7=
z\,\frac{1-\frac{1}{4}{\hat y}^2}{1+\frac{1}{4}{\hat y}^2} \ , 
\ee
in terms of which the metric takes the form
\be
ds^2_{_{S_5}}=\frac{d\hat y_i d\hat y_i }{\left(1+\frac{1}{4}{{\hat y}}^2\right)^2}
~,  \ \ \ \ ~~~~~~~
{\hat y}^2=
({{\hat y}}^4)^2+({{\hat y}}^5)^2+({{\hat y}}^6)^2+({{\hat y}}^8)^2+({{\hat y}}^9)^2 \ . \la{py}
\ee
As was mentioned in \cite{rt}, it is useful to introduce a new coordinate system 
in the Poincare patch of $AdS_5$
\be \la{klw}
ds^2_{_{AdS_5}} = 
dr^2 + e^{-2r} dx^m dx_m  = dr^2 + (dy^m + y^m dr) (dy_m + y_m dr) \ , 
\ee
where $z= e^r$ and $y^m = {x^m \over z} = e^{-r} x^m$. 
Here $m=0,1,2,3$ and the boundary signature is $(-,+,+,+)$. 
Furthermore, one may define 
\be\la{qk}
y^+ = y^0 + y^1=v\ e^w \ , \ \ \ \ \ \ \  \ \  y^- = y^0 - y^1=v\ e^{-w}\ , \ \ \ \  \ \ \ 
y^k= (y^2 ,  y^3) ~.
\ee
The choice of signs here is adapted to the special solution we are going to consider.\foot{ 
If $v, w$  are  real we are selecting the region where $y^m \geq 0$; other choices are related by analytic continuation.}
Then the $AdS_5$ metric \rf{klw} takes the form ($k=2,3$) 
\be 
ds^2_{_{AdS_5}} = dr^2  -  (dv + v dr)^2   + v^2 dw^2 
+ \sum_{k=2}^3 (dy_k + y_k dr)^2   \ . 
\label{ads_redef_metric}
\ee
We see that  shifts of $r,w$ are  linear isometries, and 
 that $v$ is an apparent time-like coordinate. 

\bs

To identify the isometry direction $\phi$ on $S^5$ 
and thus to account for  the 
presence of the chemical potential $\nu$
% it is useful to 
we perform the following coordinate transformation  $(i=4,5,6, 7)$
\be
&&{\hat z}^i=\frac{{ y}^i}{1+\frac{1}{4}y^2}\ , 
~~~~~~  \ \ \  ~~~~~~~~~~ y^2= \sum_{i=4}^7(y^i)^2\ , \cr
&&{\hat z}^8=
\frac{1-\frac{1}{4}y^2}
     {1+\frac{1}{4}y^2}\, \cos\phi\equiv Y\cos\phi \  , 
~~~~~~~~~~~~~
{\hat z}^9=
\frac{1-\frac{1}{4}y^2}
     {1+\frac{1}{4}y^2}\,\sin\phi\equiv Y\sin\phi\ . 
\ee
Then the  $S^5$ metric takes the form 
\be
ds^2_{_{S^5}}=
\left(\frac{1-\frac{1}{4}y^2} {1+\frac{1}{4}y^2}\right)^2d\phi^2
+\frac{1}{\left(1+\frac{1}{4}y^2\right)^2}\sum_{i=4}^7(dy^i)^2 \ . \la{sds}
\ee
These coordinates are well-suited for following the computational strategy described 
in sections \ref{ZQ} and \ref{WSfq}.

\subsection{Classical value of the  action}

Transforming the spinning folded string solution \rf{solpp}
to the above  coordinates and taking into account the 
rescaling of  the world sheet coordinates by $\mu$ leads to
the following form of the solution 
\be
{\bar r}=n_1\cdot{\boldsymbol{\sigma}}+\ln \sqrt 2
~,~~~~~~~~~
{\bar w}=n_2\cdot{\boldsymbol{\sigma}}
~,~~~~~~
{\bar v}=\frac{1}{\sqrt{2}}
~, ~~~~~~~~
{\bar \phi}=i\, {\rmh}\cdot {\boldsymbol{\sigma}}
~, ~~~~~~~
\la{pu}\\
n_1=(-\hkappa,1)
~,~~~~~~n_2=(\hkappa,1)\ , \ \ \ \ \  
{\rmh}=(\lnu,0)
~,~~~~~~~~
n_1\cdot n_2=-{\rmh}\cdot {\rmh}=- \lnu^2  \ , 
\label{sol}
\ee
where ${\boldsymbol{\sigma}}=(\tau, \sigma)$  and 
  the  relation between  $n_{1},n_2 $ and $\lnu$ is implied by 
  the Virasoro constraints.

%AT
%4pi  corrected

The classical string action evaluated on this solution is
\be
{\bar I}&=&\frac{\sqrt{\lambda}}{4\pi}\Big[n_1\cdot n_1-\frac{1}{2} (n_1+n_2)\cdot(n_1-n_2)-\rmh^2)\Big] \ V_2\cr
&=&\frac{\sqrt{\lambda}}{4\pi}(n_1\cdot n_1-\rmh^2)V_2
= \frac{\sqrt{\lambda}}{2\pi}\  V_2 ~.
\ee
Thus, as already mentioned in  \rf{hl}, we find that
the tree-level term in the effective action \rf{gas} is 
\be\la{kq}
{\cal F}_0=1~~.
\ee

\subsection{Fluctuation Lagrangian}

Since the solution \rf{sol} has linear coordinate dependence along 
some isometry directions of the metric \rf{ads_redef_metric}, the straightforward expansion 
$\Phi={\bar\Phi}+{\tilde \Phi}
$
($\Phi$ is a generic bosonic field with 
background 
value  ${\bar \Phi}$ and ${\tilde \Phi}$ is its fluctuation) 
of the bosonic  Lagrangian around this solution leads immediately to a
fluctuation  action 
with constant coefficients. 

The propagator of the $AdS_5$ modes is found to be 
% (cf. \ci{rt})
\be
K_{B, AdS_5}^{-1}(p)&=&
\begin{pmatrix}
\frac{1}{2p^2}+\frac{(n_2\cdot p)^2-(n_1\cdot p)^2}{p^2\Sigm_B[p] }
& \frac{in_2\cdot p}{\Sigm_B[p]}+ \frac{2 n_1\cdot p\; n_2\cdot p}{p^2\Sigm_B[p]}
& \frac{in_2\cdot p}{\sqrt{2}\Sigm_B[p]}-\frac{p^2}{2\sqrt{2}\Sigm_B[p]} 
& 0_{1\times 2 }
\cr
-\frac{in_2\cdot p}{\Sigm_B[p]}+ \frac{2 n_1\cdot p\; n_2\cdot p}{p^2\Sigm_B[p]}
& \frac{1}{p^2}-\frac{2 (n_2\cdot p)^2}{p^2\Sigm_B[p]}
& -\frac{i n_2\cdot p}{\sqrt{2} \Sigm_B[p]}
& 0_{1\times 2 }
\cr
-\frac{in_2\cdot p}{\sqrt{2}\Sigm_B[p]}-\frac{p^2}{2\sqrt{2}\Sigm_B[p]} 
& \frac{i n_2\cdot p}{\sqrt{2}\Sigm_B[p]}
& -\frac{p^2}{4\Sigm_B[p]}
& 0_{1\times 2 }
\cr
0_{2\times 1}
& 0_{2\times 1}
& 0_{2\times 1}
& \frac{\id_{2\times 2}}{2(p^2+n_1\cdot n_1)}
\end{pmatrix}
\cr
\Sigm_B[p]&=&(p^2)^2+2[(n_1\cdot p)^2+(n_2\cdot p)^2] \ \ . \vphantom{{}^{\Big|}}
\ee
One can show that  $K_B^{-1}$ describes
four massive and one massless propagating modes. 
%AT
The massless mode and a similar massless mode in $S^5$ direction mentioned below
reflect a residual conformal symmetry  in the conformal gauge and 
decouple from the non-trivial part of the computation (their constant contribution to the 1-loop 
partition function cancels against that of conformal gauge ghosts). 

The denominator $\Sigm_B$  of 
this  bosonic propagator is not 2d Lorentz invariant,
 but becomes invariant  in the limit  ${\lnu}\rightarrow 0$, 
\be 
(\Sigm_B[p])_{{\lnu}\rightarrow 0}\ \  \mapsto\ \  p^2(p^2+4)~~.
\ee
There is thus a 
 massless mode which is 
  lifted at nonzero $\lnu$; its mass arises from an imbalance between the space-like and time-like components of
   the momentum, suggesting
that in the quantum theory this 
mode will  not lead to  the  logarithmic thresholds $\sim \ln \lnu$. 

The propagator of the $S^5$ modes is substantially simpler,
\be
K_{B, S^5}^{-1}(p)=
\begin{pmatrix}
\frac{1}{2p^2} & 0_{1\times 4}             \cr
0_{4\times 1}  & \frac{\id_{4\times 4}}{2(p^2+\lnu{}^2)}
\end{pmatrix}\ . 
\ee
It describes  one massless mode and four massive modes. 
The latter become  massless   in the $\lnu \to 0 $ limit, so that 
one should  
% entitled 
 expect that the effective action should  contain $\lnu$-dependent 
threshold terms  $\lnu{}^n \ln^m \lnu$ with $n$ and $m$ depending on 
 loop order.

\

The expansion of the fermionic part of the superstring action 
is somewhat more involved. 
Finding a convenient (i.e. constant-coefficient) form of the fluctuation  action 
requires a sequence of rotations as well as a rescaling. This is so because 
in the coordinates \rf{action} neither $r$, $w$ or $\phi$ correspond
 to linear isometries. Inspecting the action \rf{action}, it is not 
 hard to see that the relevant transformations  ($\theta \to \psi$) are of the form
\be
&&\theta=e^{-{\check r}}{\cal M}_{01}(\check w){\cal M}_{89}(\check\phi)\ \psi\ , 
\cr 
&&{\cal M}_{89}(\check\phi)=\cos\frac{\check\phi}{2}
-\sin\frac{\check\phi}{2}\Gamma_8\Gamma_9
\ , \ \ \ \ \ \ \ \ \ \ \ \
{\cal M}_{01}(\check w)=\cosh \frac{\check w}{2} +\sinh \frac{\check w}{2}\,
 \Gamma_0\Gamma_1~~.\la{tras}
\ee
There is a certain amount of freedom in the choice of parameters for these 
transformations as long as they contain the background values of the fields, i.e.
\be\la{gq}
{\check r}={\bar r}+\alpha{\tilde r}
~,~~~~~~~
{\check w}={\bar w}+\beta{\tilde w}
~,~~~~~~~
{\check \phi}={\bar \phi}+\gamma{\tilde \phi}
\ee
where  $\alpha, \beta$ and $\gamma$ are some numbers. 
The values of these parameters 
are irrelevant at the  1-loop order. 
Choosing 
$\alpha=\beta=\gamma=1$ leads to a relatively simple Lagrangian. However, despite 
this simplicity  the pattern of  cancellations at higher loops is 
somewhat obscure. It turns out that $\alpha=1$, $\beta=\gamma=0$ makes it most transparent. The Lagrangian for $\alpha=1$ and arbitrary $\beta$ and $\gamma$
is ($i=2,3$; $q=4,5,6,7$):
\be
{\cal L}_F&=&
4\ \epsilon^{ab} \ {\bar\theta} \bigg[ 
\big( v {\partial_a r }+\partial_a v)(\cosh(w-{\check w})\Gamma_0
+\sinh(w-{\check w})\Gamma_1)\cr
&&~~~~~~~~~~~
+vdw(\sinh(w-{\check w})\Gamma_0 +\cosh(w-{\check w})\Gamma_1)\cr
&&~~~~~~~~~~~
+\big({\partial_a r } { y}^i+  \partial_a {y}^i \big)  \Gamma_i \ 
+\big(
{\partial_a r } {\hat z}^u +  \partial_a {\hat z}^u
\big) \Gamma_u \vphantom{\bigg|} \cr
&&~~~~~~~~~~~
+ \  
\big(
{\partial_a r } Y+   \partial_a  Y\big) (\cos(\phi-{\check\phi})\Gamma_8 +\sin(\phi-{\check\phi})\Gamma_9)\cr
&&~~~~~~~~~~~
+\  Y\partial_a{\phi}\ (-\sin(\phi-{\check\phi})\Gamma_8+ \cos(\phi-{\check\phi})\Gamma_9) \bigg]  \partial_b\theta \cr
&+&
2\ \epsilon^{ab} \ {\bar\theta} \bigg[ \big(
{\partial_a r } { y}^i+  \partial_a {y}^i \big) \partial_b{\check w} 
\Gamma_i \Gamma_0\Gamma_1\ 
+
\big({\partial_a r } {\hat z}^q +  \partial_a {\hat z}^q
\big)\partial_b{\check w} \Gamma_q\Gamma_0\Gamma_1 \cr
&&~~~~~~~~~~~
+ \  
\big(
{\partial_a r }  Y+   \partial_a  Y\big)\partial_b{\check w} (\cos(\phi-{\check\phi})\Gamma_8 +\sin(\phi-{\check\phi})\Gamma_9)\Gamma_0\Gamma_1\cr
&&~~~~~~~~~~~
+\  Y\partial_a{\phi}\partial_b{\bar w}\ (-\sin(\phi-{\check\phi})\Gamma_8+ \cos(\phi-{\check\phi})\Gamma_9)\Gamma_0\Gamma_1  \bigg] \theta  
\cr
&-&
2\ \epsilon^{ab} \ {\bar\theta} \bigg[ 
\big( v {\partial_a r }+\partial_a v)(\cosh(w-{\check w})\Gamma_0
+\sinh(w-{\check w})\Gamma_1)\ \partial_b{\check\phi}\Gamma_8\Gamma_9\cr
&&~~~~~~~~~~~
+vdw(\sinh(w-{\check w})\Gamma_0 +\cosh(w-{\check w})\Gamma_1)\ \partial_b{\check\phi}\ \Gamma_8\Gamma_9\cr
&&~~~~~~~~~~~
+\big(
{\partial_a r } { y}^i+   \partial_a {y}^i
\big)\ \partial_b{\check\phi}\  \Gamma_i \Gamma_8\Gamma_9\ 
+\big(
{\partial_a r } {\hat z}^u+   \partial_a {\hat z}^u
\big)\ \partial_b{\check\phi}\ \Gamma_u \Gamma_8\Gamma_9  \bigg]\theta  \ . \la{gqq}
\ee
The quadratic part of this action is independent of the choice of 
$\alpha$, $\beta$ and $\gamma$. Extracting and inverting it is somewhat tedious 
but algorithmic; the 
fermion propagator turns out to be ($n \times m \equiv  \ep^{ab} n_a m_b$)
\be
K_F^{-1}(p)&=&\frac{1}{\Sigm_F[p]}\Big[2\sqrt{2}\,i\,
\Big(n_1\times p\; (\Gamma_0+\sqrt{2}\Gamma_8) + n_2\times p\;\Gamma_1\Big)-4\,{\rmh}\times p \;\Gamma_8\cr
&&-i\sqrt{2}\;{\rmh}\times n_1(\sqrt{2}\;\Gamma_{019} + \Gamma_{089}+\Gamma_{189})+2\,n_1\times n_2\;\Gamma_{018}\Big]\cr
&&~~~~~~~~~~~~~~\times\Big[2\, 
%%({\rm h}\cdot{\rm h} - n_1\cdot n_1) 
(4 p^2 + 2 n_1\cdot n_1 -  n_1\cdot n_2)\id
%\cr
%&&~~~~
-8{\rmh}\cdot p\;
      %(-n_1\cdot n_1+{\rmh}\cdot{\rm h})
      \Gamma_{0189}\Big]{\cal C}^{-1}\ , 
\cr
\Sigm_F[p]&=&8
% (n_1\cdot n_2+n_1\cdot n_1)
%({\rm h}\cdot{\rm h}-n_1\cdot n_1)
\left[16 ({\rmh}\cdot {p})^2 +
 ({\rmh}\cdot {\rmh} +2 n_1\cdot n_1+ 4 {p}^2 )^2\right]
\vphantom{{}^{\Big|}}~.
\ee
Here the matrix structure was organized to emphasize the strategy used to 
construct it.  ${\cal C}$ is the charge conjugation matrix (see \ci{rtt,rt}). 
Note that similarly to the denominator of the bosonic propagator, $\Sigm_F$ 
%does not exhibit two-dimensional Lorentz invariance but  it
 becomes 2d  Lorentz-invariant in the limit  ${\rmh}\rightarrow 0$, i.e. $\lnu \to 0$:
\be
\Sigm_F[p]_{\lnu\to 0} \  \mapsto\  32(2p^2+1)^2 \ . 
\ee

\section{Quantum corrections to the effective action}

Let us now use  the  above fluctuation action to compute
the 2-loop correction to the superstring partition function or the effective action, 
$
\G= \G_0 + \G_1 + \G_2+ ..., \ \  \G_n = O({ 1 \ov (\sql)^n})$.
In the next section we will  extract from it the  corresponding correction to the 
string energy as a function of the spins in the scaling limit.

\subsection{One loop}

The one-loop partition function for the $(S,J)$ solution in the scaling limit
was computed in \cite{ftt}. 
Let us  review it here in the framework and notation set up in sections \ref{WSfq} and \ref{ZQVir}. As discussed there, we will need ${\cal F}_1(\lnu)$ in \rf{gas}, \rf{uo}
 to find the ${\cal O}({1\ov \sqrt{\lambda}})$ correction to
  the space-time energy of the folded spinning string.

The one-loop correction to the effective action is the difference between the logarithms of the determinants of the bosonic and the fermionic kinetic operators. Directly computing these determinants is quite tedious, especially for the fermions.
Taking the  product of the denominators that appear in the bosonic and fermionic propagators
leads, however, to  a useful factorization of these determinants. 

Up to a constant term which simply counts the number of   bosonic degrees 
of freedom and cancels
 against its fermionic counterpart, the bosonic contribution is 
\be
{\Gamma_{1 B}}&=&\hkappa^2  V_2\int_0^\infty  \frac{dp}{2\pi} \,\Big[
 \sqrt{p^2+2+2\sqrt{1+\hu^2p^2}}+\sqrt{p^2+2-2\sqrt{1+ \hu^2p^2}}\cr
&&~~~~~~~~~~~~~~~~~~~~~~~~+\ 2\sqrt{p^2+2-\hu^2}
+4\sqrt{p^2+\hu^2}~~\Big]\\
&=&\hkappa^2 V_2\int_0^\infty \frac{dp}{2\pi}\Big[
\sqrt{4\hu^2 +\big(p+\sqrt{p^2+4-4\hu^2}\big)^2}+2\sqrt{p^2+2-\hu^2}
+4\sqrt{p^2+\hu^2}~\Big]
\nonumber
\ee
where (see \rf{yt})
\be\la{kid}
\hkappa=\sqrt{1+\lnu{}^2}\ , \ \ \ \ \ \ \ \ \  \ \ \
\hu\equiv \frac{\lnu}{\hkappa}=\frac{\lnu}{\sqrt{1+\lnu{}^2}}~~.
% \ \ \ \ \ \ \ \ 
%\hkappa=\sqrt{1+\lnu{}^2}
\label{uofl}
\ee
The 
fermionic fluctuation spectrum is given by the solutions of the following equation:
\be
16 ({\rmh}\cdot {p})^2 + ({\rmh}\cdot {\rmh} +2 n_1\cdot n_1+ 4 {p}^2 )^2=0\ , 
\ee
i.e.
\be
\omega_\pm(p) =\sqrt{p^2 + \hkappa^2}\pm \frac{\lnu}{2}~~.
\ee
Factorizing $\Sigm_F[p]$ it is easy to find that 
 up to a constant term which cancels against its 
 bosonic counterpart, the fermion contribution to the 1-loop effective action is
\be
\Gamma_{1F}&=&-8  \hkappa^2
V_2\int_0^\infty \frac{dp}{2\pi}\,\sqrt{p^2 + 1}   \ . 
\ee
Combining the bosonic and fermionic contributions to the effective action
reproduces the result of \cite{ftt}
\be
{\Gamma_1}&=&\!\!
\hkappa^2
V_2\int_0^\infty \frac{dp}{2\pi}\Big[\sqrt{4\hu^2 
+\big(p+\sqrt{p^2+4-4\hu^2}\big)^2}+2\sqrt{p^2+2-\hu^2}\cr
&&~~~~~~~~~~~~~~~~~~~~~~~~
+4\sqrt{p^2+\hu^2}-8\sqrt{p^2 + 1}~~\Big]\\
&=&\!\!
-{\hkappa^2 \ov 2 \pi}  V_2\Big[1-\hu^2-\sqrt{1-\hu^2}+
(2-\hu^2)\ln[\sqrt{2-\hu^2}(1+\sqrt{1-\hu^2})]+2 \hu^2\ln \hu\Big]\no
\ee
Using the expressions for $\hkappa(\lnu)$  and  $u(\lnu)$ in \rf{uofl} 
this  leads to
the one-loop term in ${\cal F}(\lnu)$ in \rf{uo}
%\be
%{\cal F}_1(\lnu)=-{\hkappa(\lnu)^2}
%\Big[1-u^2-\sqrt{1-u^2}+(2-u^2)\ln[\sqrt{2-u^2}(1+\sqrt{1-u^2})]+2 u^2\ln u\Big]~~.
%\ee
\be
{\cal F}_1(\lnu)&=&-
%\frac{\hkappa(\lnu)^2}{1+\lnu{}^2}
1+\sqrt{1+\lnu{}^2}+ 2(1+\lnu{}^2)\ln(1+\lnu{}^2)- 2\,\lnu{}^2\ln\lnu \cr
&&~~~~~~~~~~~~~~~~~~~~~~~
- \ (2+\lnu{}^2)\ln\Big(\sqrt{2+\lnu{}^2}\ \big(1+\sqrt{1+\lnu{}^2}\ \big)\Big)
~~.
\la{vb}
\ee

\subsection{Two loops}

As was discussed in detail  in our earlier 
work \cite{rtt,rt}, the two loop order   is the first 
order at which it is crucial to choose 
 explicitly  a consistent regularization of the 
world sheet superstring theory. 
Among the required 
features of such a regularization should be  the  preservation 
of the   $\kappa$-symmetry of the classical action. Since this symmetry
 is intrinsically two-dimensional (having  self-dual parameters),
 the standard
  dimensional regularization is not among the consistent choices. 
  
  A scheme advocated 
in \cite{rtt, rt} 
%which preserves $\kappa$ symmetry 
is based on  doing  all algebraic
manipulations in $d=2$ and then continue the final two-dimensional 
momentum integrand to $d=2-2\epsilon$. 
While such a prescription  may be a source of confusion in  a generic  quantum 
field theories (e.g., 
 it may  not  be completely clear what the  ``final integrand'' actually 
means) in computing the partition function  of 
 an 
 anomaly-free (in particular,  finite) 
 2-dimensional theory with the 
 conformal invariance spontaneously broken by the classical solution 
there should be no ambiguity in its implementation. 

Indeed, it is expected that the integrands corresponding to  
 all potential logarithmically divergent contributions 
 cancel out before the actual integration.
 % and this may be  easily exposed. 
Only power-like divergent integrals (with no softer singularities
hidden under the leading one) may remain and they 
can be analytically (e.g.  dimensionally) regularized away.
In an explicit  cutoff regularization such power-like divergences would 
cancel against the contribution
 of the path integral measure 
 and of the nonpropagating $\kappa$-symmetry ghosts. 
An important consequence of this regularization scheme
 is that the BMN point-like string remains a BPS state at 
 the two-loop order -- its energy is not corrected  \cite{rtt}; 
 this  would  not  happen  if  one used
 the standard  dimensional regularization.

This scheme has a number of interesting and useful features that were  already 
 observed in \cite{rt} at $J=0$. In particular, using it one finds that 
 the two-loop terms in the 
 %AAT
 partition  functions 
of the $AdS_5$ and $S^5$ bosonic sigma models vanish  when computed in our classical
background. 
 As we will see, 
this  continues to be true also  in the presence of the angular momentum on $S^5$
(i.e. for $\lnu\not=0$). In particular, there will be  no logarithmic UV divergences 
coming from bosons. A consequence of this two-loop finiteness of the bosonic sigma models is
that the fermionic contribution to the two-loop partition function must also  be 
separately finite, and, indeed,  it is.

 We shall illustrate  our computational procedure on the example 
of the $S^2$ bosonic sigma model in Appendix B.

\begin{figure}[ht]
\centerline{\includegraphics[scale=0.5]{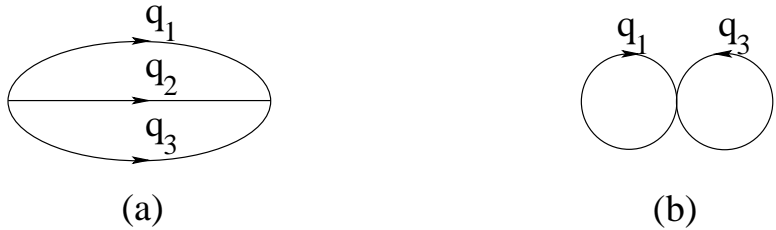}}
\caption{Topologies of  possible two-loop diagrams; each line denotes either
a bosonic or a fermionic propagator.
\label{toptwo}}
\end{figure}

The 2-loop effective action may be written as a sum of purely bosonic 
and mixed boson-fermion 
terms:
\be
\Gamma_2(\lnu)=\frac{1}{2 \pi \sqrt{\lambda}}\,V_2 \ {\cal F}_2(\lnu) \ \ , 
%\mu^2\int d^2\sigma 
%\, \left(A_B(\lnu)+A_F(\lnu)\right)
 \ \ \ \ \ \    ~~~~~~~
{\cal F}_2(\lnu)=A_B(\lnu)+A_F(\lnu)\ . 
\label{2loopgamma}
\ee
Each of them receives contributions from a ``sunset'' topology 
(figure \ref{toptwo}a, giving $A_{3B}$, $A_{3F}$) and 
a ``double-bubble'' topology (figure \ref{toptwo}b, giving  $A_{4B}$, $A_{4F}$).
Including  the appropriate combinatorial factors we have 
\be
A_B=-\frac{1}{12}A_{3B}+\frac{1}{8}A_{4B}\ , \ \ \ \ \ 
~~~~~~~~
A_F=\frac{1}{16}A_{3F}+\frac{1}{8}A_{4F}~.
\label{combinatorics}
\ee

It is not hard to produce integral representations for $A_B$ and $A_F$ using 
the Feynman rules corresponding to  the action \rf{action} expanded around the solution \rf{sol}. 
The result, however, cannot be efficiently analyzed due to the Lorentz-noninvariance 
of the denominators of  the bosonic and fermionic propagators. 
A way around this technical problem is to expand in $\hu= {\lnu\ov \hkappa(\lnu)}$. 
This is justified in light
 of the discussion below equation \rf{h_exp}. 
 %We will
 % therefore compute 
  %$A_{B,F}^{(0)}$  
%\be
%A_{B,F}(\lnu)=\hkappa^2\Big[A_{B,F}^{(0)}- \hu^2 
%\frac{\lnu{}^2}{\hkappa(\lnu)^2}
 %A_{B,F}^{(1)}+{\cal O}\left(
 %\frac{\lnu{}^4}{\hkappa(\lnu)
 %\hu^4\right)\Big]\ , \ \ \ \ \ \ \ \ \ \ \hu={\lnu\ov \hkappa(\lnu)} \ . 
%\ee
%where 
%\be \la{gah}
%A_{B,F}^{(0)} = A_{B,F}{}_{\hu=0} \ , \ \ \ \ \ \ \ \ \ \ 
%A_{B,F}^{(1)}= -({ d A_{B,F} \ov d \hu^2} ){}_{\hu=0} \ . \ee
Because the $S^5$ fluctuations become massless as $\hu\rightarrow 0$ limit,
we expect that the leading order correction has the structure
\be\la{gah}
A_{B,F}=\hkappa^2 \Big[A_{B,F}^{(0)}-{\hu}^2 
\left(a_{B,F} \ln {\hu}^2 + b_{B,F}\right)+{\cal O}({\hu}^4)\Big]~~.
\ee
We will first compute its  derivative with respect to ${\hat u}^2$  
\be
A_{B,F}^{(1)}=\frac{d}{d{\hat u}^2}  [ {\hat u}^2 \left(a_{B,F} 
\ln {\hat u}^2 + b_{B,F}\right) ] 
 = 
a_{B,F} \ln {\hat u}^2+a_{B,F}+b_{B,F}\ , 
\ee
and then integrate $A_{B,F}^{(1)}$ to determine 
$A_{B,F}$.

Let us define the integrals
\be
I{\scriptstyle \left({{a_1}\atop{m_1}}{{a_2}\atop{m_2}}{{a_3}\atop{m_3}}\right)}=\frac{1}{(2\pi)^4}
\int  \; 
\frac{d^2 k_1d^2 k_2d^2 k_3\;\; \delta^{(2)}(\sum_{i=1}^3 k_i)}
{(k_1^2+m_1^2)^{a_1}(k_2^2+m_2^2)^{a_2}(k_3^2+m_3^2)^{a_3}}\ . 
\ee
They are simple generalizations of the integrals that appeared for 
$J=0$ \cite{rt}. The higher powers in the denominator are related to 
the our expanding the 2-loop partition function in powers of 
$u$.

In terms of these integrals, the bosonic contributions are
\be
%\begin{array}{ll}
&&A_{3B}^{(0)}=
-12I{\scriptstyle \left({{1}\atop{1}}{{1}\atop{1}}{{1}\atop{\sqrt{2}}}\right)}
+6I{\scriptstyle \left({{1}\atop{\sqrt{2}}}{{1}\atop{\sqrt{2}}}\right)}\ ,  \ \ \ \ \ \ \ \ \ \ \ \ \ \ \ \
%& 
A_{4B}^{(0)}=4I{\scriptstyle \left({{1}\atop{\sqrt{2}}}{{1}\atop{\sqrt{2}}}\right)}
\\
[5pt]
&&A_{3B}^{(1)}= 
-12I{\scriptstyle \left({{1}\atop{\sqrt{2}}}{{2}\atop{\sqrt{2}}}\right)}
+3I{\scriptstyle \left({{1}\atop{\sqrt{2}}}{{1}\atop{\sqrt{2}}}\right)}
-12I{\scriptstyle \left({{1}\atop{\hu}}{{1}\atop{\hu}}\right)}\ , \\
%& \no
 &&\ \ \ \ \ \ \ \ \ \ +\ 6 I{\scriptstyle \left({{2}\atop 1}{{1}\atop 1}{{1}\atop{\sqrt{2}}}\right)}
 +  6 I{\scriptstyle \left({{1}\atop 1}{{2}\atop 1}{{1}\atop{\sqrt{2}}}\right)}
 + 12 I{\scriptstyle \left({{1}\atop 1}{{1}\atop 1}{{2}\atop{\sqrt{2}}}\right)}
 - 12 I{\scriptstyle \left({{1}\atop 1}{{1}\atop 1}{{1}\atop{\sqrt{2}}}\right)} \nonumber\\[2pt]
&& A_{4B}^{(1)}=-8I{\scriptstyle \left({{1}\atop{\sqrt{2}}}{{2}\atop{\sqrt{2}}}\right)}
   +2I{\scriptstyle \left({{1}\atop{\sqrt{2}}}{{1}\atop{\sqrt{2}}}\right)}
   -8I{\scriptstyle \left({{1}\atop{\hu}}{{1}\atop{\hu}}\right)} \ . 
 %&
%\end{array}
\ee
%
%In terms of these integrals, the bosonic contributions are
%\be
%A_{3B}^{(0)}&=&
%-12I{\scriptstyle \left({{1}\atop{1}}{{1}\atop{1}}{{1}\atop{\sqrt{2}}}\right)}
%+6I{\scriptstyle \left({{1}\atop{\sqrt{2}}}{{1}\atop{\sqrt{2}}}\right)}
%\nonumber\\[5pt]
%A_{3B}^{(1)}&=&
%-12I{\scriptstyle \left({{1}\atop{\sqrt{2}}}{{2}\atop{\sqrt{2}}}\right)}
%+3I{\scriptstyle \left({{1}\atop{\sqrt{2}}}{{1}\atop{\sqrt{2}}}\right)}
%-12I{\scriptstyle \left({{1}\atop{\lnu}}{{1}\atop{\lnu}}\right)}
%\nonumber\\[3pt]
%&+& 6 I{\scriptstyle \left({{2}\atop 1}{{1}\atop 1}{{1}\atop{\sqrt{2}}}\right)}
% +  6 I{\scriptstyle \left({{1}\atop 1}{{2}\atop 1}{{1}\atop{\sqrt{2}}}\right)}
% + 12 I{\scriptstyle \left({{1}\atop 1}{{1}\atop 1}{{2}\atop{\sqrt{2}}}\right)}
% - 12 I{\scriptstyle \left({{1}\atop 1}{{1}\atop 1}{{1}\atop{\sqrt{2}}}\right)} 
%\ee
%\be
%A_{4B}^{(0)}&=&4I{\scriptstyle \left({{1}\atop{\sqrt{2}}}{{1}\atop{\sqrt{2}}}\right)}
%\nonumber\\[5pt]
%A_{4B}^{(1)}
%&=&-8I{\scriptstyle \left({{1}\atop{\sqrt{2}}}{{2}\atop{\sqrt{2}}}\right)}
%   +2I{\scriptstyle \left({{1}\atop{\sqrt{2}}}{{1}\atop{\sqrt{2}}}\right)}
%   -8I{\scriptstyle \left({{1}\atop{\lnu}}{{1}\atop{\lnu}}\right)}
%\ee
%
Combining them as in \rf{combinatorics} immediately leads to
\be
\label{B0}
A_B^{(0)}&=&
%\frac{1}{2}M^2
I{\scriptstyle\left({{1}\atop{1}}{{1}\atop{1}}{{1}\atop{\sqrt{2}}}\right)}\ , 
\\[5pt]
A_B^{(1)}
&=&- \frac{1}{2} I{\scriptstyle \left({{2}\atop 1}{{1}\atop 1}{{1}\atop{\sqrt{2}}}\right)}
   - \frac{1}{2} I{\scriptstyle \left({{1}\atop 1}{{2}\atop 1}{{1}\atop{\sqrt{2}}}\right)}
   - I{\scriptstyle \left({{1}\atop 1}{{1}\atop 1}{{2}\atop{\sqrt{2}}}\right)}
   + I{\scriptstyle \left({{1}\atop 1}{{1}\atop 1}{{1}\atop{\sqrt{2}}}\right)}\ . 
\label{B1}
\ee

It is important to note that, despite our expansion in $\lnu$,
the result is still IR-convergent. This may seem surprising given the fact that some of the $AdS_5$ fluctuations as well as all of  the $S^5$ fluctuations become massless as 
$\lnu\rightarrow 0$.\foot{The massless $AdS_5$ fluctuation is not problematic 
because for nonzero $\lnu$ its  ``mass'' arises from a
 rotation-like imbalance between the space-like and  the 
 time-like components of momenta rather than from  a genuine mass term.} 
 The absence of 
both UV and IR divergences can be traced to the regularization scheme we 
are using. 
%AAT
Indeed, by doing the algebraic manipulations in two dimensions 
we find  that 
the full  bosonic two-loop $S^5$ contribution to the partition function is
identically zero (apart from a power divergent term, see also Appendix B).  
%Intuitively, this is due to the fact that in the absence of
% a dimensionful parameter related to the 
 %regularization scheme the $S^5$ partition 
 %function must be trivial 
 %{\it to all orders} 
% since it can only  depend  on $\lnu$ which 
%Scan be  simply scaled out of the classical action of the $S^5$ sigma model.

This ``non-regularization'' prescription is 
forced upon on us as 
the only consistent 
regularization scheme in the presence of the Green-Schwarz fermions.
It leads to the  results that are  fully consistent  with the 
expectation that the bosonic $AdS_5$ and $S^5$ sigma models 
 are embedded in a two-dimensional conformal field theory, i.e. 
it can be viewed  as  a 
 part of the definition of the quantum superstring theory.

\   

The fermion contribution is somewhat more involved.
Depending on the precise choice of the parameters $\beta$ and $\gamma$ in \rf{gq},\rf{gqq},
 different  diagrams
contribute differently. In particular, if the fermionic rotation contains
quantum fields then there are nontrivial cancellations between 
the sunset and double-bubble topologies.  
While we have checked that the final expression  is independent of the 
choice of $\beta$ and $\gamma$, the simplest
 way to state the result is by choosing $\beta=\gamma=0$, i.e.
to use a rotation involving only the  classical fields. 
Then the double-bubble contribution vanishes identically 
 while that of the sunset topology provides 
the entire fermionic contribution to the effective action.

An important subtlety  is that the naive $\lnu\to 0$ expansion
 of the fermionic sunset graphs 
involving two fermionic and one bosonic propagator 
leads to logarithmic 
IR divergences which appear because  the 4 ``transverse'' 
$S^5$ bosonic modes become  massless   at $\lnu=0$.

To isolate the relevant $\ln \lnu$ terms 
we shall  compute the derivative 
of the partition function with respect to $\lnu{}^2$  and then  set 
$\lnu=0$ everywhere except in the $S^5$ propagators. The result is a 
collection of logarithms and rational functions. The rational functions
are of course not trustworthy beyond ${\cal O}(\lnu{}^2)$. Moreover, 
only the constant term in the coefficient of $\ln \lnu$ (which leads 
to $\lnu{}^2\ln \lnu$ term in the final answer) does not receive further 
corrections.

Following this strategy, the relevant contributions are found to be 
\be
%\begin{array}{ll}
&&A_{3F}^{(0)}
=-32I{\scriptstyle \left({{1}\atop{1}}{{1}\atop{1}}{{1}\atop{\sqrt{2}}}
\right)}\ , \ \ \ \ \ \ 
~~~~~~~~~~
A_{4F}^{(0)}=0\ , \ \ \ \ \ \ \ ~~~~~ A_{4F}^{(1)}=0 \ , 
\\[5pt]
&&A_{3F}^{(1)}
=
+\ %%R 48
16 I{\scriptstyle \left({{1}\atop{1}}{{1}\atop{1}}{{1}\atop{\sqrt{2}}}\right)}
+16I{\scriptstyle \left({{1}\atop 1}{{1}\atop 1}{{1}\atop{2}}\right)}
+32I{\scriptstyle \left({{1}\atop{1}}{{1}\atop{1}}{{2}\atop{\sqrt{2}}}\right)}
%&
\no
\\[5pt]
&&~~~~~~~~
 +\ 16I{\scriptstyle \left({{1}\atop{1}}{{3}\atop{1}}{{1}\atop{\sqrt{2}}}\right)}
 + 16I{\scriptstyle \left({{3}\atop{1}}{{1}\atop{1}}{{1}\atop{\sqrt{2}}}\right)}
 - 16I{\scriptstyle \left({{1}\atop{1}}{{2}\atop{1}}{{1}\atop{\sqrt{2}}}\right)}
 - 16I{\scriptstyle \left({{2}\atop{1}}{{1}\atop{1}}{{1}\atop{\sqrt{2}}}\right)}
\no
\\[5pt]
&&~~~~~~~~
-\ 32I{\scriptstyle \left({1\atop 1}
{{1}\atop1}{{1}\atop{\hu
%\frac{\lnu}{\hkappa(\lnu)}
}}\right)} \ . 
%&
%\end{array}
\ee
%
% To restore eliminate only the first %
%
%\be
%A_{3F}^{(0)}
%&=&-32I{\scriptstyle \left({{1}\atop{1}}{{1}\atop{1}}{{1}\atop{\sqrt{2}}}\right)}
%\cr
%A_{3F}^{(1)}
%&=&
%+48I{\scriptstyle \left({{1}\atop{1}}{{1}\atop{1}}{{1}\atop{\sqrt{2}}}\right)}
%+16I{\scriptstyle \left({{1}\atop 1}{{1}\atop 1}{{1}\atop{2}}\right)}
%%\cr
%%&+&
%+32I{\scriptstyle \left({{1}\atop{1}}{{1}\atop{1}}{{2}\atop{\sqrt{2}}}\right)}
%\nonumber\\[3pt]
%&+&16I{\scriptstyle \left({{1}\atop{1}}{{3}\atop{1}}{{1}\atop{\sqrt{2}}}\right)}
% + 16I{\scriptstyle \left({{3}\atop{1}}{{1}\atop{1}}{{1}\atop{\sqrt{2}}}\right)}
%%\cr
%%&-&
% - 16I{\scriptstyle \left({{1}\atop{1}}{{2}\atop{1}}{{1}\atop{\sqrt{2}}}\right)}
% - 16I{\scriptstyle \left({{2}\atop{1}}{{1}\atop{1}}{{1}\atop{\sqrt{2}}}\right)}
%\nonumber\\[3pt]
%&-&
%32I{\scriptstyle \left({1\atop
% 1}{{1}\atop1}{{1}\atop{\frac{\lnu}{\hkappa(\lnu)}}}\right)}
%%\ee
%%\be
%\\[3pt]
%A_{4F}^{(0)}&=&0~~~~~~~~~~~~~~~A_{4F}^{(1)}=0
%\ee
%
Combining these terms  according to \rf{combinatorics} 
we are  led to the following  fermion contribution to the effective action:
\be
\label{F0}
&&A_F^{(0)}= -2I{\scriptstyle \left({{1}\atop{1}}{{1}\atop{1}}{{1}\atop{\sqrt{2}}}\right)}\ , 
\\[5pt]
&&A_F^{(1)}=%%R 3
I{\scriptstyle \left({{1}\atop{1}}{{1}\atop{1}}{{1}\atop{\sqrt{2}}}\right)}
+I{\scriptstyle \left({{1}\atop 1}{{1}\atop 1}{{1}\atop{2}}\right)}
+2I{\scriptstyle \left({{1}\atop{1}}{{1}\atop{1}}{{2}\atop{\sqrt{2}}}\right)}
\nonumber\\[3pt]
&& \ \ \ \ \ \ \ \   + \ I{\scriptstyle \left({{1}\atop{1}}{{3}\atop{1}}{{1}\atop{\sqrt{2}}}\right)}
 + I{\scriptstyle \left({{3}\atop{1}}{{1}\atop{1}}{{1}\atop{\sqrt{2}}}\right)}
 - I{\scriptstyle \left({{1}\atop{1}}{{2}\atop{1}}{{1}\atop{\sqrt{2}}}\right)}
 - I{\scriptstyle \left({{2}\atop{1}}{{1}\atop{1}}{{1}\atop{\sqrt{2}}}\right)}
\nonumber\\[3pt]
&& \ \  \ \ \ \ \ \ - \ 
2I{\scriptstyle \left({1 \atop 1}{{1}\atop 1}{{1}\atop{\hu
%\frac{\lnu}{\hkappa(\lnu)}
}}\right)}\ . 
\label{F1}
\ee
Using  \rf{B0},\rf{F0},\rf{B1},\rf{F1} and the values 
of individual integrals listed in the appendix we find that
\be
\label{finA0}
A^{(0)}=A_B^{(0)}+A_F^{(0)}&=&
%-I{\scriptstyle \left({{1}\atop{1}}{{1}\atop{1}}{{1}\atop{\sqrt{2}}}\right)}
-\frac{2}{(4\pi)^2} {\rm K} \ , 
\\[3pt]
A^{(1)}=A_B^{(1)}+A_F^{(1)}&=&
%4I{\scriptstyle \left({{1}\atop{1}}{{1}\atop{1}}{{1}\atop{\sqrt{2}}}\right)}
%+I{\scriptstyle \left({{1}\atop 1}{{1}\atop 1}{{1}\atop{2}}\right)}
%+I{\scriptstyle \left({{1}\atop{1}}{{1}\atop{1}}{{2}\atop{\sqrt{2}}}\right)}
%\nonumber\\[1pt]
%&+&I{\scriptstyle \left({{1}\atop{1}}{{3}\atop{1}}{{1}\atop{\sqrt{2}}}\right)}
% + I{\scriptstyle \left({{3}\atop{1}}{{1}\atop{1}}{{1}\atop{\sqrt{2}}}\right)}
% - \frac{3}{2}I{\scriptstyle %\left({{1}\atop{1}}{{2}\atop{1}}{{1}\atop{\sqrt{2}}}\right)}
% - \frac{3}{2}I{\scriptstyle %\left({{2}\atop{1}}{{1}\atop{1}}{{1}\atop{\sqrt{2}}}\right)}
%\nonumber\\[1pt]
%&-&
%2I{\scriptstyle \left({1 \atop 1}{{1}\atop %1}{{1}\atop{\frac{\lnu}{\hkappa(\lnu)}}}\right)}
%\nonumber\\[1pt]
%&=&
\frac{1}{(4\pi)^2}\Big( 4\ln \hu +   %R7
3{\rm K}+3\ln 2 -\frac{1}{2} 
%\frac{\lnu{}}{\hkappa(\lnu)}
\Big)~,
\label{finA1}
\ee
where ${\rm K}=0.915...$ is the Catalan's constant.

Reconstructing the two-loop effective action \rf{2loopgamma} 
implies that the first two terms in the small $\lnu$ expansion of 
${\cal F}_2(\lnu)$ in \rf{gas} are
\be
{\cal F}_2(\lnu)&=&- {\hkappa^2}\;\Big[
{{\rm K}}
+\frac{1}{2}\hu^2\big(4\ln \hu + %R7
3{\rm K}+3\ln 2-\frac{5}{2} \big)+{\cal O}(\hu^4)\Big]\ , \la{tuu}
\ee
where as 
 in \rf{uofl} we have 
  $\hkappa =\sqrt{1+\lnu{}^2}  ,\ \  \hu=\frac{\lnu}{\sqrt{1+\lnu{}^2}}$.
Though we are expanding in small $\lnu$, we have kept the full 
$\hu(\lnu)$ instead of just  its leading-order term $\lnu$  to emphasize 
that $\hu$  is the natural world-sheet   expansion parameter in the scaling limit.

\def \ttE {\bar \tE}

%%%%%%%%%%%%%%%%%%%%%%%%%%%%%%%%%%%%%%%%%%%%%%%%%%%%%%%
\section{Quantum corrections to the string energy}
%%%%%%%%%%%%%%%%%%%%%%%%%%%%%%%%%%%%%5

Having computed the effective action as a function of $\lnu$ we are now in 
position to reconstruct the difference  $E-S$ to two-loop order using 
\rf{pas},\rf{oo}--\rf{pp}.

At zero and one loops   we find 
$\tE_0$ and $\tE_1$ that reproduce the tree-level  \ci{ft1}
 and one-loop \ci{ftt} terms  in the energy of the $(S,J)$ spinning string  
 in the  scaling limit as given in \rf{uu}  and \rf{fa}.
 Using the expressions for ${\cal  F}_0$ and ${\cal F}_1$ in \rf{kq} and \rf{vb}  
 and ${\cal  F}_2$ from \rf{tuu} we can then use  the general relation \rf{pp} 
 to get the final 
  expression for the 2-loop coefficient in $E-S$ in \rf{oo}. Let us write \rf{pp} as 
 \be \la{ppt}
 {\tE}_2 =  {\ttE}_2 + \Delta {\tE}_2 \ , \ \ \ \ \ \ \ \ 
  {\ttE}_2=
 \frac{1}{\sqrt{1+\jjj^2}} \ {\cal F}_2(\jjj)  \ , \ \ \ \ \  \ \ \ \ \
 \Delta {\tE}_2 =  \ha (1  + \jjj^2)^{3/2}\,\left( \frac{d {\tE}_1(\jjj)}{d\jjj}\right)^2 \ , 
 \ee
 where ${\ttE}_2$ is the ``genuine'' 2-loop contribution coming directly from the 
 effective action  (i.e.  from the 2-loop
 graphs)  while $\Delta {\tE}_2$ is the ``one-loop'' correction 
 due  to the shift 
 of  $\jjj$  in \rf{reno}
 which arises because of  the finite renormalization \rf{pyy}
 of the relation between the $PSU(2,2|4)$ 
 charges and the parameters of the classical solution.
 
 Using  that to the leading order $\hu$ in \rf{uofl}
 can be replaced by (see \rf{vnm})
 \be
{\bu}=\frac{\jjj}{\sqrt{1+\jjj^2}}~~, 
\ee
 but keeping  as above  in \rf{tuu} the full dependence on $\bu$ in $\bar {\tE}_2$
 we get 
 \be \la{wh}
%\be E-S&=&\frac{\sqrt{\lambda}}{\pi}\ln \frac{S}{J}\left[
%{\tE}_0+\frac{1}{g}{\tE}_1+\frac{1}{g^2}{\tE}_2+\dots\right]\ , \la{ku}\\
%{\tE}_0&=&\sqrt{1+\jjj^2}\vphantom{\Big|}\la{pe} \ , \\ \la{wwq} {\tE}_1
%&=&
%-\frac{\sqrt{1+\jjj^2}}{2\pi}
%\Big[1-{\bar u}^2-\sqrt{1-{\bar u}^2}
%   +(2-{\bar u}^2)\ln[\sqrt{2-{\bar u}^2}(1+\sqrt{1-{\bar u}^2})]
%    +2 {\bar u}^2\ln {\bar u}\Big]\cr
  %  &=&-\frac{1}{\sqrt{1+\jjj^2}}\Big[\;
   % 1-\sqrt{1+\jjj^2}-2(1+\jjj^2)\ln(1+\jjj^2) +2\jjj^2\ln\jjj\\
   % &&~~~~~~~~~~~~~~~~~~~~~~~~~~~~~~~~~~~~~
   %  +(2+\jjj^2)\ln(\sqrt{2+\jjj^2}(1+\sqrt{1+\jjj^2}))
   % \Big]  
   % \nonumber  \ee
 % also   the 2-loop term  
{\ttE}_2&=&-{\sqrt{1+\jjj^2}}\;\left[\;{{\rm K}}+\frac{1}{2}\;
{\bu}^2\big( 4\ln {\bu} +  %R7
3{\rm K}+3\ln 2-\frac{5}{2} \big)
+{\cal O}({\bu}^4)\right]\ , \\
%\Delta \tE_2 
%\vphantom{\Bigg|}
%&\simeq&-\frac{1}{(4\pi)^2}\left[2{{\rm K}}+
%\jjj^2\big(7{\rm K}+3\ln 2-\frac{1}{2}+4\ln \jjj\big)
%+{\cal O}(\jjj^4)\right]+\Delta E_2\cr
\Delta {\tE}_2
%%&=&\frac{1}{(4\pi)^2}
%%\frac{\jjj^2\,\sqrt{1+\jjj^2}}{(1+\sqrt{1-{\bar u}^2})^2}\Big[-(2-{\bar %%u}^2)\sqrt{1-{\bar u}^2}+\\
%%&&~~~~~~~~~~+(1+\sqrt{1-{\bar u}^2})\left(2(2-{\bar u}^2)\ln {\bar u} 
%%+ {\bar u}^2 \ln [\sqrt{2-{\bar u}^2}(1+\sqrt{1-{\bar u}^2})]\right)\Big]^2\cr
%\cr
%&\simeq& \frac{4\jjj^2}{(4\pi)^2}\left(2\ln \jjj-\frac{1}{2}\right)^2+{\cal O}(j^4)
%%
&=&\frac{1}{2}\;\frac{\jjj^2\;(1+\jjj^2)^{-3/2}}{(1+\sqrt{1+\jjj^2})^2}\Big[
-(2+\jjj^2)+(1+\sqrt{1+\jjj^2})\Big(2(\,2+\jjj^2)\ln \jjj \no
\\
&&
~~~~~~~~~~~~~~~\ \ \ \
+\jjj^2\ln\big[\sqrt{2+\jjj^2}(1+\sqrt{1+\jjj^2})\big]
-2(1+\jjj^2)\ln(1+\jjj^2)\Big)
\Big]^2
\la{ttq}
\ee

Expanding these two contributions 
 to  ${\tE}_2$ to order $\jjj^2$ we  find 
\be
 {\ttE}_2&\simeq&-{{\rm K}}-  \;
\jjj^2\Big(2\ln \jjj  + 
\,%R4
2{\rm K}+{3 \ov 2} \ln 2-\frac{5}{4}\Big)
+{\cal O}(\jjj^4) \ , \\
\Delta{\tE}_2&\simeq& {2\jjj^2}\Big(2\ln \jjj-\frac{1}{2}\Big)^2+{\cal O}(\jjj^4)\ , 
\ee
so that finally (cf. \rf{hjk}) 
\be
&& {\tE}_2=-{\rm K}+ \,\jjj^2\Big(\,8\ln^2 \jjj-6\ln \jjj + q_{02}\Big)+{\cal O}(\jjj^4)
\ , \la{fina}\\
 && \ \ \ \ \ \ \ \ \ \ 
q_{02}= 
-  %R4 
2{\rm K} - { 3\ov 2}\ln 2 + \frac{7}{4} \ . 
\label{ina}
\ee
We observe   that the $\jjj^2\ln^2 \jjj $ and the $\jjj^2\ln \jjj $
 terms are precisely the 
same as in \rf{ggff},\rf{pred}  as predicted  in \cite{am2}.

%Remarkably,  {both} the leading $\jjj^2\ln^2 \jjj$
% {\it and}  the subleading $\jjj^2  \ln \jjj$ logarithmic
%terms in  the two-loop correction to $E-S$ \rf{oo}
%match the conjecture made in \cite{am2}. 

The  leading $\jjj^2 \ln^2 \jjj$
term originates solely from the 
contribution of the one-loop ``charge renormalization''  while 
the subleading $\jjj^2 \ln \jjj$ term receives contributions from 
 both the  genuine two-loop term
(the fermion graph  with one 
bosonic propagator from $S^5$, i.e.  with the mass proportional to $\jjj$) 
  and the one-loop ``charge renormalization''.

It was  argued  in \cite{am2} that all the fermionic
terms can be ignored in the computation of the coefficients of the leading 
$n$-loop terms $\jjj^2 \ln^n \jjj$. Our direct computation confirms this.
At the same time,  
the  coefficient of the subleading 2-loop term $\jjj^2 \ln \jjj$ 
is sensitive to 
the fermionic contributions. 
In \ci{am2}  the value of this  term  was 
predicted  by using  
the coefficient    contained in  the $\jjj^2$ part  of the 
1-loop superstring correction
 \rf{fa},\rf{dop}.\foot{The relation  between our result and the argument 
 in  \ci{am2}
may  be understood as follows.
Let us assume that we first integrate out
all fields with masses $\sim {\cal O}(1)$  as well as the modes of the
lighter fields with frequencies $\sim {\cal O}(1)$ and larger. The
result is an effective action for the $S^5$ modes with frequencies
below ${\cal O}(1)$.  Then keeping only the
2-derivative term in this effective action
 (which is the one relevant for the $\ell^2$
corrections) the coefficient of the  $O(6)$ sigma model
Lagrangian  term will be shifted,  e.g.,  by the ``heavy'' fermionic loop
contribution. The subsequent computation of the partition function
amounts to  closing  the loop for the remaining ``light'' 
 $S^5$ modes and
evaluating the momentum integrals with an ${\cal O}(1)$ UV cutoff.
This  should then  effectively produce the same contribution as
coming from the  2-loop graph with two fermionic and one $S^5$
propagator which in our case contributed to the $\ell^2 \ln \ell$
coefficient.}

% Let us  assume that   we first integrate all 
%``heavy''   modes out  getting an effective action for light $S^5$ modes
%(that means integrating out all modes up to a ``light'' scale).  
%Then   keeping only the 2-derivative term in it (which is  the one
% relevant for the $\jjj^2$
%corrections)   the coefficient of the $O(6)$ model term will 
%be shifted, e.g., 
% by the ``heavy'' fermionic loop contribution. 
 %The  subsequent computation of the 
% partition function amounts to ``closing'' the loop 
% for the light $S^5$ modes. This
% then should effectively produce the
% same  contribution  as coming from the 2-loop graph with two fermionic
%  and one $S^5$ propagator
%which in our case contributed to the $\jjj^2 \ln \jjj$ coefficient.}

\

%Let us finish with  few  comments.
Let us now comment on higher-order terms.  
An  information on the structure of the logarithmic terms at higher loops
can be inferred from  the equations \rf{q} and \rf{o} and the 
form  of the fluctuation action. 
It is plausible  that  the $S^5$ contribution to the partition function  will be  trivial 
to all orders in perturbation theory in our regularization scheme. Logarithmic
 terms can arise,  however,   only from diagrams containing
 light  $S^5$ fields and thus 
the leading one  should come from diagrams with the 
maximal  number of such fields. 
Since there is no direct coupling between the $AdS_5$ and $S^5$ fluctuations 
in the conformal gauge, such
diagrams must necessarily include  fermionic fields.
 At each loop order $n$ it is not hard to 
identify diagrams containing 
 $(n-1)$  light $S^5$ fields (for example, such is 
  an $n$-loop sunset graph with 
two fermionic propagators). 
Barring miraculous cancellations, it is then guaranteed
that the coefficient of the  $\ln^{n-1} \jj$ term 
in the
 energy  should receive genuine $n$-loop corrections.\foot{
The leading IR singularity in the relevant sunset graph is
$\displaystyle{\lim_{u\rightarrow 0}\int d^2 x 
\left(K_0(u x)\right)^{n-1}\left(K_0(x)\right)^2\propto
\ln^{n-1} u}$~ (cf. \rf{wh}).
} 

At 3-loop order  one can  see directly that  diagrams with  more that 
two  $S^5$ fluctuations have,  in fact, 
 two fields of mass $\lnu$ and all other fields of mass zero
  and thus  cannot produce higher than  $\ln^2 \jj $ contribution. A similar 
analysis can be carried out at the 4-loop order. It is therefore  natural 
 to expect that 
the leading $n$-loop logarithms $b_n(\jj)\ln^n \jj$ 
will  be  ``induced'' by the ``charge renormalization'' procedure  from 
the 1-loop
partition function, 
while the  genuine $n$-loop corrections are relevant only 
for  the first subleading term $b_{n-1}(\jj)\ln^{n-1} \jj$.

%\foot{The fact that  
 %there are no $\jj^n \ln^2 \jj$  terms coming from 
%the  2-loop graphs 
%has its origin  in our  computation scheme 
%in which the full \adss  superstring  theory 
%is finite and the $S^5$  sigma model partition function 
%is two-loop finite.} 

%% (as was already explained in the  $J=0$  case  in \ci{rt}). 
%In this natural  scheme  there are  then no $\ln^2$  UV  
%or IR (for $J \to 0$)
% divergences   at {\it two} 
 %(and apparently higher)
 % loops.

%-- Computing the $\jj^4$ term in the effective action is technically doable, but it seems %highly unlikely it will add anything new

%\

\def \co {{\cal O}}

\section{Resumming the logarithms: towards 
%understanding
interpolation to gauge-theory results}

To summarize,  combining the small $\jjj$ expansions  of the tree-level
\rf{uu}, one-loop \rf{dop} and two-loop \rf{fina}
terms in \rf{expa},\rf{fef} we get 
\be 
E-S= {\sql \ov \pi}\bigg\{  f(\l)  &+& \big[ \ q_{00}\ \jjj^2 +   \co(\jjj^4)\big]
  \cr
&+& { 1 \ov \sql} \Big[ \  
 \jjj^2 \ (q_{11}  \ln \jjj  +  q_{01} ) +   \co(\jjj^4) \Big] \   \la{hpp}\\
&+&{ 1 \ov (\sql)^2} \Big[  
\ \jjj^2\ \big(  2 q_{11}^2 \ln^2 \jjj  + 2q_{11} q_{01} \ln \jjj + q_{02} \big)+
    \co(\jjj^4) \Big] 
 + \co({1 \ov (\sql )^3}) \bigg\}\no
 \ee
 where  $f(\l)$ is given by \rf{feg} and the coefficients 
 $q_{rn}$ are defined by (see \rf{exc})\foot{Note that, unlike the universal 
 scaling function $f(\lambda)$, the coefficients thus defined do not
have definite 
transcendentality properties. It is possible that this is an artifact 
%R/K
of the 
 small $\jjj$ expansion at large $\lambda$. } 
 \be \la{qw}
q_0(\l) &=& 
\sum_{n=0}^\infty { q_{0n} \ov (\sql)^n} 
=   \ha  + {3 \ov 2\sql }  +  {1 \ov (\sql)^2 }(
 -  %R4 
2{\rm K} - { 3\ov 2}\ln 2 + \frac{7}{4})\  + \ \co ({1 \ov (\sql) ^3}) 
\ , \\
q_1(\l) &=& \sum_{n=1}^\infty { q_{1n} \ov (\sql)^n}  =- { 2 \ov \sql } + 
\co({1 \ov (\sql)^2})
\la{lpo}\ . \ee
 Remarkably, all of the explicitly written 
leading terms in \rf{hpp} can be reproduced 
 by expanding in $1 \ov \sql$ 
  the following  expression  (see \rf{feg},
\rf{exc})
\be 
E-S= {\sql \ov \pi}\bigg\{\ f(\l) 
+  { q_0(\l)\ \jjj^2    \ov 1  - 2 q_1(\l)  \ln \jjj }  + ...  
+ \  \co(\jjj^4)\  \bigg\}
\ , \la{ruf}
\ee
%In line with the discussion  in \ci{am2} 
The  coefficient of $\ln \jjj$  in 
the denominator $1  - 2 q_{11}  \ln \jjj = 1 + { 4 \ov \sql } \ln \jjj  $
has its origin  in the value (4=6-2)  of the 1-loop $\beta$ function 
of the $S^5$ sigma-model (cf. \rf{dop}). 
The above expression  obviously  resembles the   RG running coupling, 
 in agreement  with the discussion  in \ci{am2}.\foot{Notice also that the 
 subleading 2-loop $\ln \jjj$ term in \rf{hpp} 
originates from the  product of the 1-loop logarithm in the denominator 
and the  $1\ov \sql$ term  (3/2) 
in $q_0(\l)$.}
In  fact, as  was pointed out  in \ci{am2}, the  closed expression that reproduces 
the expected  coefficients of the  first
two leading $ \jjj^2\ln^n \jjj$ and $ \jjj^2\ln^{n-1} \jjj$ terms 
at $n$-th loop order in $1 \ov \sql$   expansion  is 
\be 
E-S= {\sql \ov \pi}\bigg\{\ f(\l)  
+    { \jjj^2    \ov 1   - {2  q_{11} \ov \sql}  \ln \jjj } \Big(q_{00}   + 
{q_{01}\ov \sql} {  1 + \ln (1   - {2  q_{11} \ov \sql}  \ln \jjj )
    \ov 1   - {2  q_{11} \ov \sql}  \ln \jjj }\Big) + ...+ \  \co(\jjj^4)\  \bigg\}
\ .  \la{rqf}
\ee
This expression does not include the information about the 2-loop
coefficient $q_{02}$ in \rf{fina} we have computed here 
since, in contrast to  its 1-loop counterpart $q_{01}$,  
 it  does not influence the  coefficients of the two leading   powers of $\ln \jjj$.
%
% One may try  to guess a further generalization of \rf{rqf}
% in the direction of \rf{ruf}  which would  include  the full information about 
% the function $q_0(\l)$, e.g.,
% 
One may try  to guess a further generalization of \rf{rqf} which also 
incorporates the features of \rf{ruf}. As an illustration, an example 
of possible  expression  which encodes the complete information about
 the function $q_0(\lambda)$ is\foot{Generalizations of  \rf{rqf} and
  \rf{ruf} involving coefficients of definite 
transcendentality and recovering $q_0$ in a large $\lambda$ expansion
are also easy to construct.}
 \be 
 E-S={\sql \ov \pi}\bigg\{\ f(\l)   \  +\ \frac{\jjj^2}{1-2q_1(\lambda)\ln 
 \jjj}\sum^\infty_{n= 0}\frac{q_{0n}}{(\sqrt{\lambda})^n}
\Big(\frac{1+\ln [1-2q_1(\lambda)\ln \jjj]}{1-2q_1(\lambda)\ln \jjj}\Big)^n+  \co(\jjj^4)\  \bigg\}~.
\la{kop}
\ee
A resummation of the logarithms of $\jjj$ appears to be necessary in order to interpolate 
between
the string  perturbative   expansion ($\l \gg 1$)
and the
% to
 gauge theory perturbative 
expansion $\l \ll 1$) in a similar limit. 

The  gauge-theory  expansion corresponds to fixing the value of  $j
= \sql \jjj = { \pi J \ov \ln S}$ for any  $\l$ and expanding first 
in $\l$ and then in $j$. At the  1-loop  order in the $sl(2)$ sector it was found in \ci{bgk}  
that the anomalous  dimension  scales as 
$\Delta \equiv E-S-J =  \l ( 
%R/K
a_{00}+a_{10} j + a_{30} j^3 + ...) \ln S +...$. 
Surprisingly, the absence 
of the $j^2$ term   extends to all orders  in the weak coupling expansion.
As was very recently shown  
   by a perturbative solution of $J\not=0$ generalization of the
     BES equation in the $S\to \infty$, small $j$ limit, one has  \ci{staud} 
\be \la{gai}
\Delta  = \left[   
%R/K
f(\l)+a_1 (\l) \  j + \ a_3 (\l)\ j^3 +  \ 
a_4 (\l)\ j^4\  + \ a_5 (\l)\ j^5\ + ...\right] \  \ln S\ , \ \ \ \ \ \ \ \ \  j\equiv 
 { \pi J \ov \ln S}
\ , \ee
where the functions $a_k(\l)$ are given by convergent series in $\l$ \ci{staud}.
%R

%Thus 
The corresponding weak-coupling, small $j = \sql\ \jjj$ 
 continuation of the above string theory expression  \rf{hpp}  (or  the one 
 like  in \rf{kop})
should be expected to reproduce the absence  of  all terms of the type 
$j^k \ln^n j$  as well as  the complete $j^2$  term in the energy. 
The  appearance of the logarithmic $\jjj^k \ln^n \jjj$ terms in \rf{hpp} 
should therefore be an artifact of the string perturbative expansion.

The string expression \rf{hpp} was obtained by assuming that 
$\lambda \gg 1$  for fixed $\jjj= {j \ov \sql}$ while in the gauge theory
expansion one assumes that $j=\sql \jjj$ is fixed and expands
at small $\lambda$.  Thus, to compare the  gauge and the string theory 
results we are supposed to start  with the exact form of the string 
(strong-coupling) expansion,  fix $j$ and  start decreasing $\l$.
While the  expansion in \rf{hpp} is singular in this limit, 
the resummed  expression  \rf{ruf}  or \rf{rqf} 
leads to  an apparent suppression of the terms proportional to $j^2$ 
present in \rf{hpp}:  instead of $\jjj^2$ we get 
 $ \frac{\jjj^2 \ln^n \jjj}{\ln^{n+1}\lambda}$ terms which 
vanish for small 't~Hooft coupling.\foot{The unnatural 't~Hooft coupling 
dependence also points in a  direction of a required   resummation.}

%
%The string expression \rf{hpp} was obtained by assuming that 
%\lambda \gg 1$  for fixed $\jjj= {j \ov \sql}$. 
%To compare to gauge theory we are supposed to start  with the exact form of the 
%strong-coupling expansion, fix $j$ and  start decreasing $\l$.
%That effectively corresponds to taking 
%$\jjj$ {\it large}. 
%While the expression expansionin \rf{hpp} is singular in this limit, the 
%resummed expression  \rf{ruf}  or \rf{rqf}  
%leads to a an apparent
%suppression of the unwelcome $\jjj^2\sim j^2 $ terms present in \rf{hpp}:
%we get  instead $ \jjj^2\ov \ln \jjj$ or $ \jjj^2\ov \ln^2 \jjj$ terms. 
%

This gives an indication that a resummation  of the string perturbative expansion 
may allow one to smoothly connect it with the gauge-theory expansion \rf{gai}.
For example, the expression like \rf{kop}  may produce a 
function $ j^2\;k(\lambda, \ln j)$ which  vanishes in the small $\lambda$ limit.
To systematically address this issue one needs, in fact,  to 
find the exact form of the string result to all orders in $\jjj$.

In particular, it would be important to find the exact 
%in 
$\jjj$-dependence
%form 
of the 2-loop contribution  in \rf{fef}. 
It is easy to see that   the     ${1 \ov (\sql)^2} \jjj^2 \ln^2 \jjj$   term  in 
\rf{fef}, \rf{fina}
 is   part of a more general
 term
 \be\la{tak}
{\tE}_2 = \  b_2(\jjj)\  \ln^2 \jjj +...  \ , \ \ \ \ \ \ \ \ \ \ \ \ \ \ \ \ 
b_2(\jjj)={2\jjj^2 (2+\jj^2)^2\ov  (1+\jjj^2)^{3/2}}\ . 
\ee
Here $b_2(\jj)$
is completely  determined  by the  1-loop calculation, i.e. by \rf{fa}.
While for small $\jjj$ the  function $b_2$ 
 scales as $\jjj^2$, its   small $\lambda$,  fixed $j=\sql\jjj$ 
asymptotics (relevant for a discussion of an interpolation to gauge theory) 
 is $\sim j^3$. 
  
  It would be interesting  to  find a similar exact form of the subleading 
 $\jjj$-dependent   terms in ${\tE}_2$ in  \rf{fef}
 to determine  how its $\jjj^2$  behavior discussed above 
  may change in the gauge theory limit. 
%
% It would be interesting  to  find a similar exact form of the subleading 
% $\jjj$-dependent   terms in ${\tE}_2$ in  \rf{fef}
% to find how its $\jjj^2$  behaviour discussed above 
%  may change in the  large $\jjj$ limit.
%  Note that the  exact tree-level \rf{uuuu} and the 1-loop
% \rf{fa} terms  contain $\jjj^2$ terms in their expansion at small $\jjj$ 
% but their large $\lambda$ expansion  starts  with $j$ for ${\tE}_0$ and 
% $ j^{-1}$ for  ${\tE}_1$. 
% We expect a similar  negative power asymptotics for ${\tE}_2$ at  
% $\jjj \gg 1$.
%
 Note that the  exact tree-level \rf{uuuu} and the 1-loop
\rf{fa} terms  contain $\jjj^2$ terms in their expansion at small $\jjj$ 
but their expansion in the gauge theory regime
starts  with $j$ for ${\tE}_0$ and $ j^{-2}$ for  ${\tE}_1$. 
We expect similar  negative-power asymptotics for ${\tE}_2$ at small $\lambda$ for fixed $j=\sql\jjj$.

\medskip
Finally, let us recall  that 
a  consequence of the strong coupling 
perturbative 
solution of the BES equation in    \ci{bkk} 
is that all the coefficients in  the inverse
 string tension expansion are negative and 
grow factorially.  
%Notice  that the 
%leading  $\lnu^2 \ln \lnu $ correction to the 2-loop 
%partition function \rf{tuu} has the same sign as the  Catalan's constant 
%term.
% suggesting that the
 %inclusion of the spin in the $S^5$ direction does not improve the convergence
 % properties of strong coupling perturbation theory. 
%However,  the 
%sign of the $\lnu^2 \ln \lnu $ 
%correction to the { energy}  \rf{fina} is opposite to that of 
%the Catalan's  K term. 
%While that result  is valid only in the small $\lnu$ regime, 
%the fact that for $\lnu\ne 0$ the energy contains a positive 
%(and potentially large) term 
%follows in general 
% from the equation \rf{pp}.
% 
Notice,  however, that in the  equation \rf{pp} the inclusion of the  $\jj$ dependence
corrects the ${\cal F}_2$ term  (which contains the complete 2-loop 
contribution to the universal scaling function)
by a positive (and potentially large) contribution. It is  then 
 tempting to speculate  that 
 a natural definition of the non-Borel-summable  \ci{bkk} series expansion  
for  the  scaling function  (cusp 
anomalous dimension) at strong coupling should be to start with  
a more general  $\lnu\ne 0$ case, resum the series 
 and  then consider the $\lnu\to 0$ limit.

\bigskip

%%%%%%%%%%%%%%%%%%%%%%%%%%%%%%%%%%%%%%%%%%%%%%%%%%%%%%%
\section*{Acknowledgments }
%%%%%%%%%%%%%%%%%%%%%%%%%%%

We are  grateful to F. Alday, A. Belitsky, 
N. Dorey,  S. Frolov, G. Korchemsky, J. Maldacena, A. Polyakov, 
A. Rej, M. Shifman,  M. Staudacher and K. Zarembo 
for very useful discussions
on various related issues.
 R.R.  acknowledges the support of the National Science Foundation under grant
 PHY-0608114  and of a Sloan Research Fellowship. 
 A.A.T. acknowledges the support of the EU-RTN network grant 
 MRTN-CT-2004-005104, the INTAS 03-51-6346 grant
  and the RS~Wolfson award. 
 Part of this work was done while we participated in
 the programme "``Strong Fields, Integrability and Strings''" at 
 the Isaac Newton Institute in Cambridge, U.K.

\bigskip

\renewcommand{\theequation}{A.\arabic{equation}}
\renewcommand{\thesection}{A}
 \setcounter{equation}{0}
\setcounter{section}{1} \setcounter{subsection}{0}

\section*{Appendix A:   Some 2-loop integrals}

Here  we list the integrals relevant for deriving  the 
equations \rf{finA0} and \rf{finA1}:
\be
\label{int1}
I{\scriptstyle \left({{1}\atop{1}}{{1}\atop{1}}{{1}\atop{2}}\right)}&=&
\frac{2\ln 2}{(4\pi)^2}
\\
\label{int2}
I{\scriptstyle \left({{1}\atop{1}}{{1}\atop{1}}{{1}\atop{\sqrt{2}}}\right)}&=&
\frac{2{\rm K}}{(4\pi)^2}
\\
\label{int3}
I{\scriptstyle \left({{1}\atop{1}}{{1}\atop{1}}{{2}\atop{\sqrt{2}}}\right)}&=&
\frac{1}{(4\pi)^2}\,\frac{\ln 2}{2}
\\
\label{int4}
I{\scriptstyle \left({{1}\atop{1}}{{2}\atop{1}}{{1}\atop{\sqrt{2}}}\right)}&=&
\frac{1}{(4\pi)^2}\left({\rm K}-\frac{1}{2}\ln 2\right)
\\
\label{int5}
I{\scriptstyle \left({{1}\atop{1}}{{3}\atop{1}}{{1}\atop{\sqrt{2}}}\right)}&=&
\frac{1}{(4\pi)^2}\left({\rm K}-\frac{1}{4}-\frac{1}{2}\ln 2\right)
\\
\label{int6}
I{\scriptstyle \left({1 \atop 1}{{1}\atop 1}{{1}\atop{{\sqrt{\alpha}}}}\right)}&=&
\frac{1}{(4\pi)^2} \frac{1}{\sqrt{\alpha(\alpha-4)}}\Bigg[~
\ln\frac{2-\sqrt{\alpha(\alpha-4)}-\alpha}{2+\sqrt{\alpha(\alpha-4)}-\alpha}\,\ln\alpha\\
&&~~~~~~~~~~~~~~~~~~~~~~~\;
+2{\rm Li}_2\left(\frac{2\sqrt{\alpha}}{\sqrt{\alpha}-\sqrt{\alpha-4}}\right)
-2{\rm Li}_2\left(\frac{2\sqrt{\alpha}}{\sqrt{\alpha}+\sqrt{\alpha-4}}\right)\Bigg]\cr
&=&-\frac{\ln\alpha}{(4\pi)^2}+ ...
%{\rm (\; not~relevant\,)}
\label{int6exp}
\ee
where the ellipsis stands for terms that are not relevant for us here. 
K is the Catalan's constant. 

In recovering \rf{int1} and \rf{int2} from \rf{int6} one should be careful in
 identifying the correct branch for the dilogarithms (e.g. by requiring that 
the result is real).

\renewcommand{\theequation}{B.\arabic{equation}}
\renewcommand{\thesection}{B}
 \setcounter{equation}{0}
\setcounter{section}{1} \setcounter{subsection}{0}

\section*{Appendix B:    2-loop effective action  of $S^2$ sigma model}

To illustrate the discussion  of our regularisation prescription in section 4.2 
here 
we shall discuss a simple example: the 
 computation of the 2-loop  correction to 
the partition function of
 the $S^2$  sigma model.

Let us start with the Lagrangian (in our string-theory context 
one may set $g^{-2}={\sqrt{\lambda} \ov 2\pi}$)
\be
{\cal L}=\frac{1}{2g^2}\left[\;(\partial\theta)^2+\cos^2\theta(\partial\phi)^2\right]
\ee
We shall choose the  Euclidean signature and expand in the standard way around a solution
with a linear profile in the isometric direction $\phi$:
\be
{\bar\theta}={\tilde\theta}~, \ \ \ \ ~~~~~~~~{\bar\phi}=i\nu\tau+{\tilde\phi}~~.
\ee
Ignoring the total derivative term  gives 
\be
{\cal L}&=&\frac{1}{2g^2}\left[\;(\partial{\tilde\theta})^2
+(1-{\tilde\theta}^2+\frac{1}{3}{\tilde\theta}^4)(\;-\nu^2
+2i\nu\partial_0{\tilde\phi}+(\partial{\tilde\phi})^2)\right]
\cr
&=&\frac{1}{2g^2}\left[\;(\,\partial{\tilde\theta})^2
+\nu^2{\tilde\theta}^2+(\partial{\tilde\phi})^2
-2i\nu\partial_0{\tilde\phi}\,{\tilde\theta}^2
-{\tilde\theta}^2(\partial{\tilde\phi})^2
-\frac{1}{3}\nu^2{\tilde\theta}^4\right]~~.
\ee
The propagators then are 
\be
K^{-1}_\theta=\frac{1}{p^2+\nu^2}\  ,\ \ \ \ \ \ 
~~~~~~~~
K^{-1}_\phi=\frac{1}{p^2}~~.
\ee
The 1-loop effective action is  given by ($\Lambda$ is an UV  cutoff) 
%regularization and after regularizing the massless mode and removing 
%Sthat regulator first)
\be
\Gamma_1=\ha \int\frac{d^2p}{(2\pi)^2}\;\big[\;\ln(p^2+\nu^2)+\ln p^2\;\big]=
\frac{\nu^2 }{8\pi}\Big(1-\ln\frac{\nu^2}{\Lambda^2}\Big)~~.
\ee
As in section 4.2, the 2-loop effective action receives contributions 
from diagrams with topologies shown in figure 1
\be
\Gamma_2=g^2\int d^2 p d^2 q~\left[-\frac{1}{12}A_3+\frac{1}{8}(A_4+A_4')\right]~~.
\ee
The sunset diagram (figure $1$a) contributes:
\be
A_3=-12\nu^2\frac{(p+q)_0\,(p+q)_0}{(p^2+\nu^2)(q^2+\nu^2)(p+q)^2} \ , 
\label{App:A3}
\ee
while the double bubble diagram (figure $2$b) with a $\theta^4$ vertex gives:
\be
A_4=-\frac{4\nu^2}{(p^2+\nu^2)(q^2+\nu^2)}\ . 
\label{App:A4}
\ee
Finally, the double bubble diagram (figure $2$b) 
with a $(\partial\phi)^2\theta^2$ vertex
yields a quadratically-divergent contribution 
\be
A_4'=2\frac{p^2}{p^2(q^2+\nu^2)}=\frac{2}{q^2+\nu^2}\ ,
\ee
which we shall 
ignore (it is cancelled by the measure contribution, cf. \ci{hone}).

Let us  now compare the evaluation of  the integrals of 
(\ref{App:A3}) and (\ref{App:A4}) 
in dimensional regularization and in the regularization 
used in our computation in section 4.

%\noindent
 In dimensional regularization we continue the momentum integrals to 
$d=2-2\epsilon$ dimensions from the outset. On Lorentz-invariance grounds 
we can therefore write
\be
\int
\frac{d^dp\,d^d q}{(2\pi)^{2d}}
\frac{(p+q)_a\,(p+q)_b}{(p^2+\nu^2)(q^2+\nu^2)(p+q)^2}
=\frac{1}{d}
\int
\frac{d^dp\,d^d q}{(2\pi)^{2d}}
\frac{\delta_{ab}}{(p^2+\nu^2)(q^2+\nu^2)}~~.
\ee
Taking the 00-component of this  tensor relation  one finds 
that the 2-loop effective action is given by 
\be
\Gamma_2=g^2\frac{d-2}{2d}\int
\frac{d^dp\,d^d q}{(2\pi)^{2d}\mu^{4\epsilon}}
\frac{1}{(p^2+\nu^2)(q^2+\nu^2)}\ , 
\ee
which upon evaluation leads to a nonvanishing and divergent answer:
the  $d-2$ factor cancels one of the two $1\ov d-2$ poles 
from the standard tadpole integral (this is what happens in the
 computation of the 2-loop
sigma model beta-function \ci{fri}).

\

 Within  our regularization prescription
 we stay in two dimensions. We can use the same Lorentz-invariance
  argument as above to write
\be
\int
\frac{d^2p\,d^2 q}{(2\pi)^{4}}
\frac{(p+q)_a\,(p+q)_b}{(p^2+\nu^2)(q^2+\nu^2)(p+q)^2}
=\frac{1}{2}
\int
\frac{d^2p\,d^2 q}{(2\pi)^{4}}
\frac{\delta_{ab}}{(p^2+\nu^2)(q^2+\nu^2)}
\ee
and then project onto the $00$-component.
Alternatively, we may simply notice that the denominator 
$(p^2+\nu^2)(q^2+\nu^2)(p+q)^2$ is invariant under the simultaneous 
transformation
\be
p_0\leftrightarrow p_1~, \ \ \ \ ~~~~~~~~~ q_0\leftrightarrow q_1~~.
\ee
Then symmetrizing $A_3$ under this  transformation we get 
\be
A_3&=&-12\nu^2\int
\frac{d^2p\,d^2 q}{(2\pi)^{4}}
\frac{(p+q)_0\,(p+q)_0}{(p^2+\nu^2)(q^2+\nu^2)(p+q)^2}\cr
&=&-6\nu^2\int
\frac{d^2p\,d^2 q}{(2\pi)^{4}}
\frac{(p+q)_0\,(p+q)_0+(p+q)_1\,(p+q)_1}{(p^2+\nu^2)(q^2+\nu^2)(p+q)^2}\cr
&=&-6\nu^2\int
\frac{d^2p\,d^2 q}{(2\pi)^{4}}
\frac{1}{(p^2+\nu^2)(q^2+\nu^2)}
\ee
Combining this with \rf{App:A4}
 leads to the  trivial result for the  2-loop effective action
\be
\Gamma=g^2\left(\frac{1}{2}-\frac{1}{2}\right)\int
\frac{d^2p\,d^2 q}{(2\pi)^{4}}
\frac{1}{(p^2+\nu^2)(q^2+\nu^2)}=0~~.
\ee
Thus,  in our regularization 
prescription the 2-loop term in the 
free energy of the  $S^2$ sigma model is identically zero.

%NNEW
In the 2d supersymmetric version of this computation 
in dimensional regularization
the fermions  produce a similar $(d-2) [I(d)]^2 \sim { 1 \ov d-2} + ...$ 
divergent contribution   that cancels the bosonic one, 
in agreement with the vanishing of the 2-loop $\beta$-function 
contribution in the 
supersymmetric $O(n)$ model.
In our prescription  the fermionic contribution is separately 
finite and vanishing, so the  total result for the 2-loop 
contribution to the effective action is the same as in the dimensional
regularization.\foot{The genuine  finite 
 2-loop contribution to
the free energy  of the supersymmetric $O(n)$
 model (which apparently was not directly  computed in the past)
 thus vanishes. 
 The vanishing of this  finite part  did not  play a
  role in the comparison of the perturbative
 expansion and the exact solution in  \cite{EH}. 
 In the bosonic case \ci{has} 
   the  apparent 2-loop
 contribution to  the free energy (proportional to the 2-loop beta-function
coefficient $\beta_2$ which vanishes in the supersymmetric case)
arises solely from expressing the 1-loop free energy in
 terms of the 2-loop running coupling.}

% While the coupling constant
%  of the GS superstring sigma model  we considered here
%  is not running, one may  interpret  the agreement  with the $O(6)$ model
% we have
%  found as being due to  the role of the term in the $O(6)$  free energy
% with  the 2-loop beta function coefficient
%  being played by the genuine 2-loop logarithmic dependence
%  of the partition function (cf. eq. \rf{tuu}).

\end{document}